\pdfoutput=1

\documentclass[
	amsmath,
	amssymb,
	a4paper,
	aip,		
	jcp,		
	reprint,	
	fleqn,
	showpacs,
	floatfix
]{revtex4-1}

\usepackage[T1]{fontenc}
\usepackage{lmodern}
\usepackage{microtype}

\usepackage{amssymb}

\usepackage{color}

\usepackage{tabularx}
\usepackage{booktabs}
\usepackage{multirow}

\usepackage{subfig}
\usepackage[
	format=hang,
	indention=-1.0cm,
	justification=RaggedRight,
	font=small
]{caption}

\usepackage{graphicx} 
\usepackage{dcolumn} 
\usepackage{bm} 
\usepackage[version=3]{mhchem}
\usepackage{stackrel}

\newcommand{\myTitle}{Glucans monomer-exchange dynamics as an open chemical network}

\usepackage{hyperref}

\newcommand{\myName}{Riccardo Rao}

\newcommand{\myAffiliation}{Complex Systems and Statistical Mechanics, Physics and Materials Science Research Unit, University of Luxembourg, L-1511 Luxembourg, Luxembourg}

\newcommand{\myAdvisor}{Massimiliano Esposito}

\newcommand{\myCollab}{David Lacoste}
\newcommand{\myCollabAff}{Laboratoire de Physico-Chimie Th\'eorique, UMR CNRS Gulliver 7083, ESPCI - 10 rue Vauquelin, F-75231 Paris, France}

\newcommand{\der}[2]{\frac{\mathrm{d}{#1}}{\mathrm{d}{#2}}}

\newcommand{\at}[2]{\left.{#1}\right|_{#2}}

\DeclareMathOperator{\coker}{coker}

\newcommand{\numberset}{\mathbb} 
\newcommand{\N}{\numberset{N}} 
\newcommand{\R}{\numberset{R}} 

\definecolor{webgreen}{rgb}{0,.5,0}
\definecolor{webbrown}{rgb}{.6,0,0}
\definecolor{grigio}{rgb}{.85,.85,.85} 
\definecolor{RoyalBlue}{rgb}{0.0, 0.14, 0.4}

\hypersetup{%
    colorlinks=true, linktocpage=true, pdfstartpage=1, pdfstartview=FitV,%
    breaklinks=true, pdfpagemode=UseNone, pageanchor=true, pdfpagemode=UseOutlines,%
    plainpages=false, bookmarksnumbered, bookmarksopen=true, bookmarksopenlevel=1,%
    hypertexnames=true, pdfhighlight=/O,
    urlcolor=webbrown, linkcolor=RoyalBlue, citecolor=webgreen, 
    pdftitle={\myTitle},%
    pdfauthor={\textcopyright\ \myName},%
    pdfsubject={},%
    pdfkeywords={},%
    pdfcreator={pdfLaTeX},%
    pdfproducer={LaTeX REVTeX}%
}

\begin{document}

\title{\myTitle}

\author{\myName}
\affiliation{\myAffiliation}
\author{\myCollab}
\affiliation{\myCollabAff}
\author{\myAdvisor}
\affiliation{\myAffiliation}
\date{\today. Published in \emph{J.~Chem.~Phys.} DOI:~\href{http://dx.doi.org/10.1063/1.4938009}{10.1063/1.4938009}}

\begin{abstract}
	We describe the oligosaccharides-exchange dynamics performed by so-called D-enzymes on polysaccharides.
	To mimic physiological conditions, we treat this process as an open chemical network by assuming some of the polymer concentrations fixed (chemostatting). 
	We show that three different long-time behaviors may ensue: equilibrium states, nonequilibrium steady states, and continuous growth states.  
	We dynamically and thermodynamically characterize these states and emphasize the crucial role of conservation laws in identifying the chemostatting conditions inducing them.
\end{abstract}


\maketitle

\section{Introduction}

Biological systems use large and branched chains of basic sugars, called polysaccharides, to store energy \cite{*[][{, pp.~244--246.}] {nelson08}}.
\emph{Glucans} such as \emph{glycogen} and starch are polysaccharides whose building blocks are \emph{D-glucose} monosaccharides.
Despite the apparent simplicity of their constituents, their metabolism involves several chemical steps, each performed by a specific set of enzymes \cite{ball03}.
Interestingly, some of these catalysts lack specificity regarding the reaction they catalyze or the substrates they act on
\footnote{
	Enzymes' lack of specificity is nowadays getting increased attention.
	The most notorious examples are enzymes involved in genetic processes and which are unspecific because they need to act on different substrates, e.g. polymerases and synthetases acting on the four nucleo-bases \cite{*[][{, \S~4.5.}] {bialek12}}.
	More recently, enzymes involved in metabolic processes have also been shown to lack specificity \cite{linster13}. 
	These observations have led to various theoretical studies on the thermodynamical cost of error correction \cite{hopfield74, ninio75, bennett79, andrieux08:copolymerization, bel10, murugan12, sartori13, rao15:proofreading}.
}.
An example is provided by \emph{(1$\rightarrow$4)-alpha-D-glucans} \cite{jones69,takaha99,colleoni99,ball03} (EC 2.4.1.25), also called \emph{D-enzymes}, which act on pairs of glucans regardless of their size \cite{kartal11}.
	Specifically, D-enzymes catalyze the transfer of groups of glycosyl residues from a donor glucan to an acceptor glucan \cite{jones69,takaha99}.
	Experimental evidences highlight the presence of bonds between glycosyl residues which are not cleaved by D-enzymes \cite{jones69}---at least not over physiological time scales \cite{kartal11}.
	These bonds are called \emph{forbidden linkages} \cite{jones69}.
	In this way, D-enzymes transfer segments of glucan chains containing one or more forbidden linkages, and the transfer of segments containing one forbidden linkage are the most probable \cite{jones69}.
	Also, each glucan chain is characterized by a reducing-end glucose which is not transferred by D-enzymes \cite{jones69,kartal11}.
	Hence, glucans made of just the reducing end can act only as acceptor in the transfer.

Qualitatively, D-enzymes process medium-size glucans by disproportionating them into unit-size and big-size glucans \cite{takaha99}.
Since their transfers reactions are neutral energetically \cite{goldberg91,kartal11}, entropy is the main driving force in this system. 
In closed conditions, this system evolves towards an equilibrium state maximizing the entropy \cite{kartal11,lahiri15}.

In this paper we consider a simplified kinetic description of the D-enzyme's action on glucans, which we treat as a chemical network.
Since metabolic processes should be thought of as part of an open system continuously fed from the environment, we mimic these 
physiological conditions by introducing \emph{chemostats} (i.e. species whose concentrations are kept constant by the environment).
Our goal is to characterize the dynamical and thermodynamical implications of treating the action of the D-enzymes on glucans as an open chemical network.  
In the framework of deterministic chemical networks endowed with \emph{mass action kinetics}, we prove that chemostatting can induce three different types of long-time behaviors: equilibrium, non-equilibrium steady state, and continuous growth.
	The equilibrium state corresponds to the stationary concentration distribution in which the concentration currents along each reaction pathway vanishes (detailed balance property \cite{*[See for instance: ][{, \S~9.4.}] {kondepudi14}}).
	Non-equilibrium steady states refer to stationary distributions violating detailed balance.
	Hence, contrary to equilibrium states, a continuous and steady flow of mass circulates across the network.
	Finally, the continuous growth regime we observed corresponds to a non-stationary state characterized by continuous and steady flow of mass entering the network, and resulting in its continuous growth.
We emphasize the dynamical and thermodynamical role of conservation laws and emergent cycles in identifying the chemostatting conditions leading to these states.
We are thus able to confirm the general relation between the number of chemostatted species and the number of independent thermodynamical forces---or \emph{affinities}---found in Ref. \cite{polettini14}.
Despite the simplicity of our description, the closed system results found in Ref. \cite{kartal11} are reproduced and the qualitative disproportionating behavior of D-enzymes \cite{takaha99} is captured by our (chemostatted) open system description.

The plan of the paper is as follows: in sec.~\ref{sec:description} the kinetic model is established and the related rate equation description for the concentration of polysaccharides is introduced.
In sec.~\ref{sec:ssc} the chemostatting conditions leading to non-equilibrium steady states rather than equilibrium ones are found.
For this purpose, both the conservation laws of the dynamics and the emergent cycles of the network are analyzed.
The dissipation of the non-equilibrium steady state is also studied.
The network's conservation laws identified in sec.~\ref{sec:cl} are used in sec.~\ref{sec:ss} to derive the steady-state concentration distributions for different numbers of chemostats.
The explosive asymptotic behavior is described in sec.~\ref{sec:growth}.
Conclusions are drawn in sec.~\ref{sec:conc}.
Some technical derivations and proofs are provided in the appendices.

\section{The Kinetic Model}
\label{sec:description}

\begin{figure}[t]
	\centering
	\includegraphics[width=.45\textwidth]{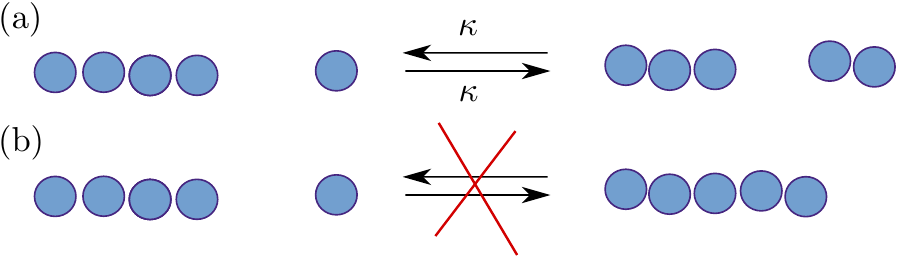}
	\caption{
		(a) The typical monomer-exchange reaction describing the action of D-enzymes on glucan chains.
		(b) The attachment of free monomers to other species is not allowed.
	}
	\label{fig:interaction}
\end{figure}

	The action of D-enzymes is modeled as follows (see also Fig.~\ref{fig:interaction}).
	Glucans are treated as polymers whose monomers represent single transferable segments.
	Hence, each glucan is identified by its number of monomers, or equivalently by its monomeric mass.
	The enzymatic steps performed by the D-enzymes in order to achieve the transfer are not explicitly described---they are coarse-grained---, and we describe the interaction between two polymers of mass $n$ and $m$ as a \emph{mass-exchange process} \cite{*[][{, ch.~5.}] {krapivsky10}}:
	\begin{multline}
		(n) + (m) \, \overset{\kappa_{nm}}{\longrightarrow} \, (n + 1) + (m - 1), \\ \text{for } n \ge 1,\, m \ge 2,
		\label{reaction}
	\end{multline}
	where $\kappa_{nm}$ denotes the related coarse-grained rate constant.
	Transfers of oligosaccharides longer than one monomeric unit are less probable \cite{jones69}, and are not considered in our description.
	We take into account the presence of non-transferable units by imposing the size of the donor glucan $(m)$ to be greater than one \cite{jones69,kartal11}.

Let us note that each reaction is reversible because the backward path is already included in \eqref{reaction} (it is realized by replacing $n \rightarrow m - 1$ and $m \rightarrow n + 1$ in the above expression).
Furthermore, the constraint on the minimal size of the donor molecules imposes that $m \geq 2$.
Since we describe the glucans as linear polymers, and since D-enzymes do not discriminate the size of the polymers, we assume a constant kernel for the reactions: $\kappa_{mn} = \kappa$, $\forall n\geq 1, \forall m\geq 2$.
This assumption is based on the evidence that the free-enthalpy release resulting from any reaction is almost vanishing \cite{goldberg91,kartal11}.
Indeed, for any bond cleaved, a new one of the same kind will be formed.

Assuming a large and well stirred pool of interacting polymers, the evolution of the system is well described by reaction rate equations \cite{*[][{, ch.~5.}] {krapivsky10}}.
According to this mean-field description, the molar concentration of polymers of mass $k$ at time $t$, $Z^k = Z^k (t)$, satisfies the following first order differential equations
\begin{equation}
	\dot{Z}^k = \frac{1}{2} \sum_{ \substack{ n \ge 1 \\ m \ge 2 } } \nabla^k_{nm} \underbrace{\left( J^{+nm} - J^{-nm} \right)}_{{}\equiv J^{nm}}, \,\, \text{for } k \ge 1 .
	\label{eq:rate}
\end{equation}
The $\frac{1}{2}$ factor in front of the summation takes into account that summing over all $n\geq 1$ and $m \geq 2$ includes every reaction pathway twice \footnote{Indeed, the symmetry of the reaction scheme \eqref{reaction} is inherited by the stoichiometric matrix, which satisfies $\nabla^{k}_{nm} = - \nabla^{k}_{m-1 \, n+1}$}.
$\nabla^k_{nm}$ represents the element of the \emph{stoichiometric matrix} related to the species of mass $k$ and to the reaction involving an acceptor and a donor polymer of mass $n$ and $m$, respectively.
The reaction scheme in \eqref{reaction} implies that
\begin{equation}
	\nabla^k_{nm} = \delta^k_{n+1} + \delta^k_{m-1} - \delta^k_{n} - \delta^k_{m} ,
	\label{expr:stoichiometric}
\end{equation}
where $\delta_{i}^{j}$ represents the Kronecker delta.
Assuming a \emph{mass action kinetics}, the forward (denoted by $+$) and the backward fluxes ($-$) can be written as:
\begin{equation}
	\begin{aligned}
		J^{+nm} & = \kappa Z^{n}Z^{m} \, ,& J^{-nm} & = \kappa Z^{n+1}Z^{m-1} \, ,
	\end{aligned}
	\label{expr:mal}
\end{equation}
where $Z^{n}$, denotes the concentrations of the polymers of size $n$.
To simplify the following discussion, we will use the Einstein summation notation: upper indexed quantities represent vectors, lower indexed ones covectors, and repeated indexes implies the summation over all the allowed values for those indexes ($1\le n \le n_{\mathrm{max}}$ and $2\le m \le m_{\mathrm{max}}$, or $1\le k \le k_{\mathrm{max}}$, where $n_{\mathrm{max}}$, $m_{\mathrm{max}}$ and $k_{\mathrm{max}}$ are finite in closed systems but infinite in open ones).
To avoid confusion, exponents will always act on parenthesis (\emph{e.g.} $(a)^{n} $ denotes the quantity $a$ to the power $n$).

The rate equations \eqref{eq:rate} assume the following form when the expressions for both the stoichiometric matrix \eqref{expr:stoichiometric} and the fluxes \eqref{expr:mal} are considered
\begin{equation}
	\begin{aligned}
		\dot{Z}^1 &= \kappa Z \left( Z^2 - Z^1 \right) + \kappa Z^1 Z^1 , \\
		\dot{Z}^k &= \kappa Z \left( Z^{k+1} - 2 Z^k + Z^{k-1} \right) + \\ 
		&\quad + \kappa Z^1 \left( Z^{k} - Z^{k-1} \right) , \quad \text{for } k \geq 2 ,
	\end{aligned}
	\label{eq:rate:explicit}
\end{equation}
where $Z \equiv \sum_{k=1}^{k_{\mathrm{max}}} Z^k$ denotes the total concentration. 
The second term in the right hand side of \eqref{eq:rate:explicit} arises from the constraint that the donor species cannot be monomers \footnote{We refer to eq.~5.99 in Ref. \cite{*[][{, ch.~5.}] {krapivsky10}} for the monomers-exchange processes without constraints} (see Fig.~\ref{fig:interaction}b).

\begin{figure}[t]
	\centering
	\includegraphics[width=.45\textwidth]{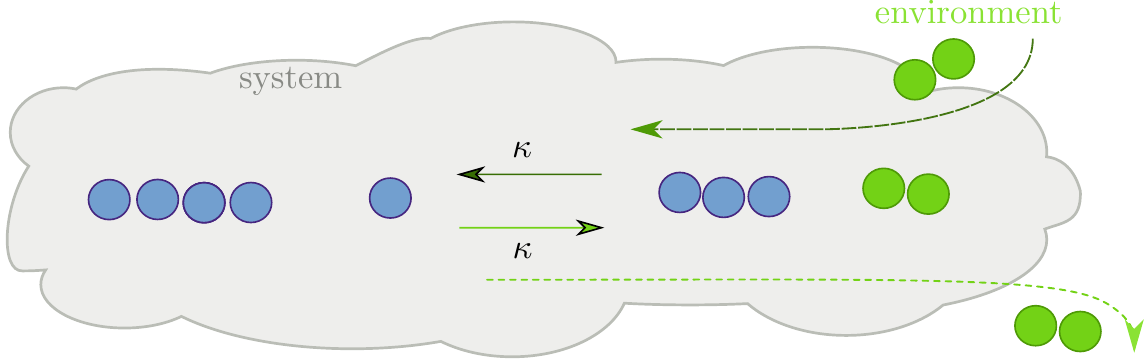}
	\caption{
		Pictorial representation of a reaction involving a chemostat.
		When a reaction produces a chemostat (here a dimer), the environment extracts one molecule of this species from the system (dotted light green reaction).
		On the other hand, when a chemostat reacts, a new molecule is injected into the system (dashed dark green reaction).
	}
	\label{fig:chemos}
\end{figure}

To model the open system we now assume that the environment keeps the concentrations of some species constant by refilling the consumed ones and eliminating the produced ones, see Fig.~\ref{fig:chemos}.
We call these species \emph{chemostats}\footnote{{
	To be precise, chemostats represent reservoirs of particles to whom the system is connected.
	However, in order to simplify the nomenclature we refer to the chemical species exchanged by the reservoirs as chemostat, as well
}}
and {we denote them with} the indices $k_{\mathrm{y}} \in \Omega_{\mathrm{Y}}$, where $\Omega_{\mathrm{Y}} \subset \N$ represents a subset of all species.
The remaining (variable) species are explicitly denoted by $k_{\mathrm{x}}$.

By definition, the chemostats' concentrations must remain constant, $\dot{Z}^{k_{\mathrm{y}}} = 0$.
The rate of chemostatted molecules consumed by the reactions in the network must therefore be balanced by the rate of chemostatted molecules injected/rejected from the system.
The rate of injection/rejection of the $k_{\mathrm{y}}$-th chemostat is quantified by the external currents~\cite{polettini14}, whose expression is
\begin{equation}
	\begin{aligned}
		& I^{k_{\mathrm{y}}} = \frac{1}{2} \nabla^{k_{\mathrm{y}}}_{nm} \left( J^{+nm} - J^{-nm} \right) = \\ 
		& = \kappa Z \left( Z^2 - Z^1 \right) + \kappa Z^1 Z^1 & & \text{if } k_{\mathrm{y}} = 1 \\
		& = \kappa Z \left( Z^{k_{\mathrm{y}}+1} - 2 Z^{k_{\mathrm{y}}} + Z^{k_{\mathrm{y}}-1} \right) \\
			&\quad + \kappa Z^1 \left( Z^{k_{\mathrm{y}}} - Z^{k_{\mathrm{y}}-1} \right) & & \text{if } k_{\mathrm{y}} \ge 2 .
	\end{aligned}
	\label{eq:chemos}
\end{equation}

\section{Steady States: conservation laws, cycles and dissipation}
\label{sec:ssc}

Three different types of long-time behaviors have been identified for our kinetic model: equilibrium, non-equilibrium steady state and continuous growth.
We start by focusing on the chemostatting conditions leading to equilibrium or non-equilibrium steady states.
The existence and uniqueness of the steady state is currently \emph{a priori} assumed.

Closed systems always reach an equilibrium steady state \cite{schuster89} defined by $\dot{Z}_{\mathrm{eq}}^{k_{\mathrm{x}}} = 0, \forall k_{\mathrm{x}}$ and $J^{nm}_{\mathrm{eq}} = 0, \forall n,m$.
Their dynamics is constrained by \emph{conservation laws} \cite{alberty03,palsson06,polettini14}, which fully characterize the equilibrium concentration distribution.
Chemostatting generic chemical species may break these conservation laws and may create chemical forces---also called \emph{affinities}\cite{polettini14}.
The appearance of affinities is directly related to that of so-called \emph{emergent cycles}, through which the external chemical forces can act.
In finite chemical networks, if no emergent affinity arises from the chemostatting procedure, the system will always relax to a unique equilibrium state compatible with the chemostats and the non-broken conservation laws \cite{schuster89, polettini14}.
When emergent cycles---or equivalently affinities---are generated, the system may evolve towards a non-equilibrium steady state defined by $\dot{\bar{Z}}^{k_{\mathrm{x}}} = 0,\,\forall k_{\mathrm{x}}$ and $\bar{J}^{nm}\neq 0$ (non-equilibrium steady state quantities are denoted by an overbar in the text). 
In the following subsections we analyze how the closed system's conservation laws and emergent cycles are modified by the gradual increase of the number of chemostatted chemical species.
In the last subsection we relate these to the dissipation in the system.

\subsection{Conservation Laws}
\label{sec:cl}

Conservation laws denote the presence of physical quantities which are conserved during the evolution of the system, the so-called \emph{components}.
In general, they can be identified from the cokernel space of the stoichiometric matrix \cite{alberty03,palsson06,polettini14}.
Indeed, if $l_k \in \coker \nabla$, namely if $l_k \nabla^{k}_{nm} = 0$, the scalar $l_k Z^k$ is conserved
\begin{equation}
	\begin{split}
		\der{}{t} \left( l_k Z^k \right) &= l_k \dot{Z}^k \\ 
		&= \frac{1}{2} \, l_k \nabla^k_{nm} \left( J^{+nm} - J^{-nm} \right) = 0 .
	\end{split}
	\label{eq:cl}
\end{equation}

For the closed system, the equation leading to the conservation laws is $l^k_{n+1} - l^k_{n} = l^k_{m} - l^k_{m-1}, \, \text{for } 1\le n \le n_{\mathrm{max}} = k_{\mathrm{max}} - 1 \text{ and } 2 \le m \le m_{\mathrm{max}} = k_{\mathrm{max}}$.
It exhibits the following solutions: $l_{k}^{\mathrm{(1)}} = \alpha$ and $l_{k}^{\mathrm{(2)}} = \alpha \cdot k$ (where $\alpha$ is an arbitrary constant, which is taken as one when expressing the components), which correspond to the conservations of the total concentration $Z \equiv \sum_{k=1}^{k_{\mathrm{max}}} Z^k$ and the total mass $M \equiv \sum_{k=1}^{k_{\mathrm{max}}} k Z^k$, respectively.
Hence, $k_{\mathrm{max}} = M - Z + 1$.

However, when the system is opened by setting chemostats, the relevant stoichiometric matrix becomes the stoichiometric submatrix of the variable species: $\nabla^{k_\mathrm{x}}_{nm}$.
Also, $k_{\mathrm{max}} = \infty$.
No matter what the sizes of the chemostatted glucans are, neither the total concentration conservation law $l_{k_\mathrm{x}} = \alpha$ nor the total mass conservation law $l_{k_\mathrm{x}} = \alpha k_\mathrm{x}$ survives (i.e. they are not anymore elements of the cokernel space of $\nabla^{k_\mathrm{x}}_{nm}$).
We therefore say that the total mass and the total concentration are \emph{broken conservation laws}.
Nevertheless, when just one chemostat is present, $\Omega_{\mathrm{Y}} \equiv \{ k_{y} \}$, a new conservation law emerges:
\begin{equation}
	l_{k_\mathrm{x}}^{\mathrm{(3)}} = \alpha \left( k_\mathrm{x} - k_\mathrm{y} \right) .
	\label{expr:cl:residual}
\end{equation}
Hence, the system exhibits just one (net) broken conservation law.
It corresponds to the component
	\begin{equation}
		q = M - k_\mathrm{y} Z ,
		\label{expr:cq:residual}
	\end{equation}
which can assume any value in $\R$ and takes into account that the total mass can change in the system only by multiples of the chemostat mass, $k_{\mathrm{y}}$.
In presence of more than one chemostat, no conservation law survives. 

The components derived in this section---$M$ and $Z$ for the closed system and $q$ for the network with one chemostat---will be used to characterize the steady state distribution in sec.~\ref{sec:ss}.

\subsection{Emergent cycles}
\label{sec:cycles}

A cycle represents a finite set of reactions which leave the state of the network unchanged.
Algebraically they are represented as vectors $c^{nm}$ and they belong by definition to the kernel space of the stoichiometric matrix ($c^{nm} \in \ker \nabla$): $\frac{1}{2}\nabla^k_{nm}c^{nm} = 0$.

The steady-state currents satisfy $\nabla^k_{nm} \bar{J}^{nm} = 0$ and can always be written as linear combinations of cycles.
The cycle space of our polymers system is however infinite dimensional and its complete characterization is of little use.
However, in order to characterize non-equilibrium steady states only the \emph{emergent cycles}---those cycles that may appear when chemostatted species are introduced---are needed \cite{polettini14}.
Physically, they represent cyclic transformations leaving the variable species $k_{\mathrm{x}}$ unchanged, but which would change the concentrations of the chemostats $k_{\mathrm{y}}$ if they were not kept constant and contribute to the external currents.

An emergent cycle ($\gamma^{nm}$) is thus defined by
\begin{equation}
	\left\lbrace
	\begin{aligned}
		\tfrac{1}{2} \nabla^{k_{\mathrm{x}}}_{nm} \gamma^{nm} &= 0 , \\
		\tfrac{1}{2} \nabla^{k_{\mathrm{y}}}_{nm} \gamma^{nm} &= \nu_{\gamma}^{k_{\mathrm{y}}} & \neq 0 \text{ for at least one } k_{\mathrm{y}} ,
	\end{aligned}
	\right.
	\label{def:emergent}
\end{equation}
where $\left\{ \nu_{\gamma}^{k_{\mathrm{y}}} \right\}_{k_\mathrm{y} \in \Omega_\mathrm{Y}}$ denotes the amount of chemostats of mass $k_{\mathrm{y}}$ injected (minus sign) or rejected (plus sign) from the chemical network during the transformation $\gamma^{nm}$.
These quantities cannot take arbitrary values, due to the constraints imposed by the conservation laws of $\nabla^k_{nm}$.
Indeed, for any conservation law, $l_k^{(i)}$, a constraint of the following form holds
\begin{equation}
	l_{k_{\mathrm{y}}}^{(i)} \nu_{\gamma}^{k_{\mathrm{y}}} = l_{k_{\mathrm{y}}}^{(i)} \tfrac{1}{2} \nabla^{k_{\mathrm{y}}}_{nm} \gamma^{nm} = 0 .
	\label{eq:constr}
\end{equation}
Taking into account the total concentration $l_{k}^\mathrm{(1)} = \alpha$ and total mass $l_{k}^\mathrm{(2)} = \alpha k$ conservation laws, derived in sec.~\ref{sec:cl} (the emergent conservation law $l_k^{(3)}$ is a linear combination of the first two on the whole set of species indexes), we obtain the following constraints
\begin{equation}
	\left\{
	\begin{aligned}
		&\textstyle\sum_{k_{\mathrm{y}}} \nu_{\gamma}^{k_{\mathrm{y}}} = 0 \\
		&\textstyle\sum_{k_{\mathrm{y}}} k_{\mathrm{y}} \nu_{\gamma}^{k_{\mathrm{y}}} = 0 .
	\end{aligned}
	\right.
	\label{eq:constr:explicit}
\end{equation}
Non-trivial solutions of this set of equations signal the presence of emergent cycles, and thus of independent affinities, which read \cite{polettini14}
	\begin{equation}
		A_{\gamma} = \frac{1}{2} \sum_{nm} \gamma^{nm} \, \ln \prod_{k_{\mathrm{y}}} (Z^{k_{\mathrm{y}}})^{- \nabla^{nm}_{k_{\mathrm{y}}}} \, .
		\label{def:affinity}
	\end{equation}
The set of linearly independent solutions of \eqref{eq:constr:explicit} gives the number of independent emergent cycles in the chemostatted chemical network.
If we normalize this set so to have the smallest non-vanishing integer values for $\nu_\gamma^{k_\mathrm{y}}$, these values indicate the number of chemostatted species which are introduced in or rejected from the system in precisely one (emergent) cyclic transformation.

For less than three chemostats, only trivial solutions of \eqref{eq:constr:explicit} exist and therefore no emergent cycle appears.
For three chemostats, we obtain one emergent cycle characterized by the following normalized values for $\nu^{k_{\mathrm{y}}}$:
\begin{equation}
	\begin{aligned}
		\nu^{k_{\mathrm{y}1}} & = k_{\mathrm{y}3} - k_{\mathrm{y}2}, \\
		\quad \nu^{k_{\mathrm{y}2}} & = k_{\mathrm{y}1} - k_{\mathrm{y}3}, \\
		\nu^{k_{\mathrm{y}3}} & = k_{\mathrm{y}2} - k_{\mathrm{y}1} ,
	\end{aligned}
	\label{expr:nu3}
\end{equation}
where $k_{\mathrm{y}1}$, $k_{\mathrm{y}2}$ and $k_{\mathrm{y}3}$ represent the masses of the chemostats. 
For any additional chemostat we obtain an additional emergent cycle, each characterized by its value for the coefficients $\nu^{k_{\mathrm{y}}}$.

\subsection{External Currents and Dissipation}
\label{sec:dissip}

We now show that at steady state, the emergent cycles determine the external currents $\bar{I}^{k_{\mathrm{y}}}$ and the entropy production rate $\Sigma$.

We first observe that the steady-state external currents $\bar{I}^{k_{\mathrm{y}}}$ are in general linear combination of the coefficients $\nu_{\gamma_i}^{k_{\mathrm{y}}}$ and must satisfy the same constraints (eq.~\ref{eq:constr:explicit}).
Indeed, the steady-state equations in presence of chemostats
\begin{equation}
	\left\lbrace
	\begin{aligned}
		\tfrac{1}{2} \nabla^{k_{\mathrm{x}}}_{nm} \bar{J}^{nm} &= 0 , \\
		\tfrac{1}{2} \nabla^{k_{\mathrm{y}}}_{nm} \bar{J}^{nm} &= \bar{I}^{k_{\mathrm{y}}} 
	\end{aligned}
	\right.
	\label{eq:ss}
\end{equation}
are equivalent to eq.~\eqref{def:emergent}: the emergent cycles $\gamma^{nm}$ are substituted by the steady state currents $\bar{J}^{nm}$ and the coefficients $\nu^{k_\mathrm{y}}$ by the steady-state external currents $\bar{I}^{k_\mathrm{y}}$.
Thereby, if no cycle emerges due to the chemostats, the steady-state external currents $\bar{I}^{k_{\mathrm{y}}}$ are vanishing, provided that the steady state exists.
The system is then at equilibrium.

The dissipation at steady state is intimately related to the external currents \cite{polettini14}.
Indeed, the (non-negative) entropy production rate for our chemical reaction network can be written as
\begin{equation}
	\begin{aligned}
		\Sigma & \equiv \frac{1}{2} \sum_{nm} J^{nm} R \ln \frac{J^{+nm}}{J^{-nm}} = \\
		& = \underbrace{- \sum_{k_{\mathrm{x}}} \dot{Z}^{k_{\mathrm{x}}} R \ln \dfrac{Z^{k_{\mathrm{x}}}}{Z^{k_{\mathrm{x}}}_{\mathrm{eq}}}}_{\equiv \Sigma_\mathrm{X}} \underbrace{- \sum_{k_{\mathrm{y}}} I^{k_{\mathrm{y}}} R \ln \dfrac{Z^{k_{\mathrm{y}}}}{Z^{k_{\mathrm{y}}}_{\mathrm{eq}}}}_{\equiv \Sigma_\mathrm{Y}} ,
	\end{aligned}
	\label{expr:epr}
\end{equation}
where $R$ is the gas constant.
At the steady state, the internal species' contribution $\Sigma_\mathrm{X}$ always vanishes.
Hence, the dissipation is characterized by the contribution due to the chemostats $\Sigma_{\mathrm{Y}}$, which is non-vanishing if the set of steady-state external currents $\bar{I}^{k_{\mathrm{y}}}$ is also non-vanishing.
	We also mention that, the steady state entropy production can be expressed as the sum along a set of independent emergent cycles of products of affinities \eqref{def:affinity} and emergent cycle currents \cite{polettini14} $\mathcal{J}_{\gamma}$: $\bar{\Sigma} = \sum_\gamma A_{\gamma} \mathcal{J}_{\gamma}$.

Summarizing, the conservation laws provide us with both the components--- which are useful for expressing the steady state distributions---and the constraints (eq.~\ref{eq:constr:explicit}) on the emergent cycles of the network (eq.~\ref{def:emergent}).
Due to these constraints, the first emergent cycle appears in the system with three chemostats.
For any additional chemostat an additional independent cycle emerges.
Through these cycles the environment exerts chemical forces, which are generated by the chemostats concentrations.
The above-defined external currents result from these forces and characterize the dissipation.

We emphasize that the relation between the number of chemostats $s^{\mathrm{Y}}$, of net broken conservation laws $b$, and of emergent cycles $a$, is in perfect agreement with the general result obtained for finite-dimensional phase space in Ref. \cite{polettini14} stating that
\begin{equation}
	s^{\mathrm{Y}} = b + a
	\label{}
\end{equation}
These results are summarized in Tab.~\ref{tab:summary}. 

\begin{table}
	\centering
	\begin{tabular}{lccr}
		\toprule
		number of   & broken  & independent & asymptotic \\
		chemostats, $s^{\mathrm{Y}}$ & c.~laws, $b$ & affinities, $a$  & behavior  \\
		\midrule
		\multirow{1}*{0} & \multirow{1}*{0} & \multirow{1}*{0} & \multirow{1}*{ES} \\
		\multirow{1}*{1} & \multirow{1}*{1} & \multirow{1}*{0} & \multirow{1}*{ES} \\
		\multirow{1}*{2} & \multirow{1}*{2} & \multirow{1}*{0} & ES/growth \\
		\multirow{1}*{3} & \multirow{1}*{2} & \multirow{1}*{1} & NESS/growth \\
		\multirow{1}*{4} & \multirow{1}*{2} & \multirow{1}*{2} & NESS/growth \\
		\bottomrule
	\end{tabular}
	\caption{
		Summary of the behaviors of our model for different numbers of chemostats (ES stands for ``equilibrium state'' whereas NESS for ``non-equilibrium steady state'').
		The number of broken conservation laws and independent affinities are also reported.
		The growth state occurs whenever the concentration of the largest chemostat is larger than the concentration of the smallest one: ($Z^{k_{\mathrm{y\,larger}}} \ge Z^{k_{\mathrm{y}1}}$).
	}
	\label{tab:summary}
\end{table}

\section{The stationary distributions}
\label{sec:ss}

We now use the components introduced in the previous section to derive the steady-state concentration distribution for different number of chemostats.
The conditions on the chemostats' concentrations not leading to the steady state solution are also identified.

From the steady-state equations corresponding to \eqref{eq:rate:explicit} and from the equations for the external currents \eqref{eq:chemos}, we can write a general expression for the steady-state concentrations as a function of the concentration of monomers, $\bar{Z}^1$, the fraction of polymers larger than monomers, $\bar{r} \equiv 1 - {\bar{Z}^1}/{\bar{Z}}$, and the chemostats fluxes, $\bar{I}^{k_\mathrm{y}}$:
\begin{multline}
	\bar{Z}^{k} = \bar{Z}^{1} (\bar{r})^{k-1} + \\ + \sum_{k_{\mathrm{y}} \in \Omega_{\mathrm{Y}}} \frac{\bar{I}^{k_{\mathrm{y}}}}{\kappa} \, \frac{1 - ( \bar{r} )^{k - k_{\mathrm{y}}}}{1 - \bar{r}} \, \Theta \left( k - k_{\mathrm{y}} - 1 \right) ,
	\label{expr:ss}
\end{multline}
where $\Theta(\cdot)$ represents the discrete step function (we refer the reader to appendix~\ref{app:ss} for details).
Here, the number of chemostats is arbitrary, and since the external currents at steady state satisfy the same constraints as in \eqref{eq:constr:explicit}, only $s^\mathrm{Y} - 2$ of them are independent.
In the next paragraphs we will discuss in detail the above expression for zero, one, two and three chemostats, and the variables $\bar{Z}^1$, $\bar{r}$ and $\bar{I}^{k_\mathrm{y}}$ will be expressed in terms of the components and of the chemostats' concentrations.

\subsection{Closed system}
\label{sec:closed}

\begin{figure}[t]
	\centering
	\includegraphics[width=.45\textwidth]{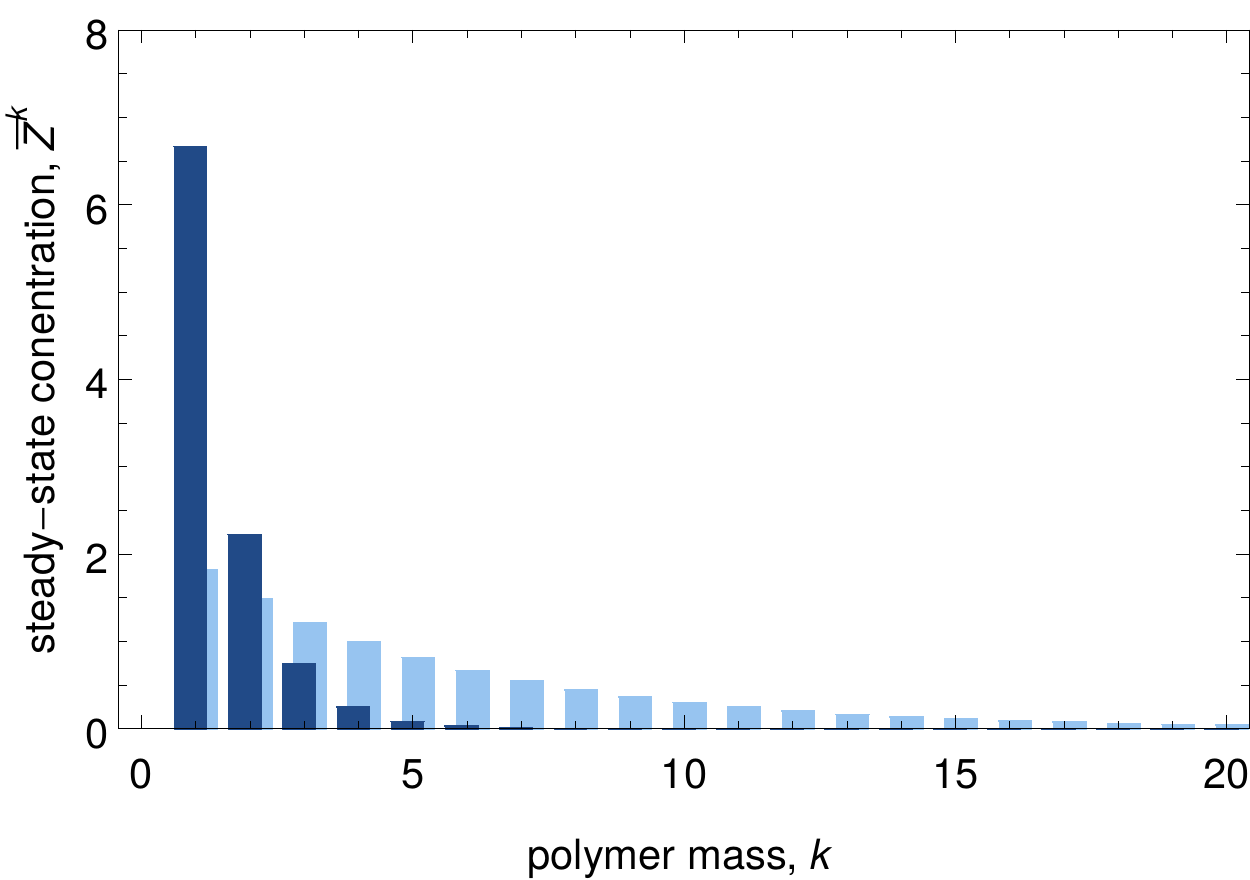}
	\caption{
		Equilibrium concentration distribution for the closed system of monomers-exchanging polymers at different values of the total concentration $Z$ and total mass $M$.
		The dark blue bar plot refers to the choice $Z = 10$ and $M = 15$, while the light blue one to $Z = 10$ and $M = 55$.
	}
	\label{fig:c:ss}
\end{figure}

As previously discussed, the closed system exhibits the following components: $Z = \sum_{k=1}^{k_{\mathrm{max}}} Z^k$ and $M = \sum_{k=1}^{k_{\mathrm{max}}} k Z^k$.
In order to express the equilibrium distribution algebraically as function of $Z$ and $M$ we consider the following limit $M \gg Z$.
In this way $k_{\mathrm{max}} \sim \infty$ and imposing $Z = \sum_{k=1}^{\infty} Z^k$ and $M = \sum_{k=1}^{\infty} k Z^k$ on the expression \eqref{expr:ss} we can write $\bar{Z}^1$ and $\bar{r}$ as functions of $Z$ and $M$.
Hence
\begin{equation}
	\bar{Z}^k = \frac{(Z)^2}{M} \left( 1 - \frac{Z}{M} \right)^{k-1} .
	\label{expr:ss:0}
\end{equation}
Fig.~\ref{fig:c:ss} shows the behavior of this distribution for different values of $Z$ and $M$.
As expected, the higher the ratio between the mass and the concentration $M \gg Z$, the broader the distribution.

\textit{Remark.}
The equilibrium distribution we obtained from our dynamical description is equivalent to the result obtained using maximum entropy approaches and is consistent with experimental observations~\cite{kartal11}.
The equivalence is inferred by comparing eq.~\eqref{expr:ss:0} with eq.~(1),~(3)~and~(4) in Ref. \cite{kartal11}.

\subsection{Open system: 1 chemostat}
\label{sec:1chemo}

Introducing a chemostat breaks the concentration and mass conservation laws, but a new one arises \eqref{expr:cl:residual}.
As a result, no affinity appears ($s^{\mathrm{Y}} = 1$, $b = 1$ and $a = 0$) and the system evolves towards an equilibrium state compatible with the chemostat concentration $Z^{k_{\mathrm{y}}}$ and the value of the component $q$ \eqref{expr:cq:residual} (the steady-state external current vanishes, $\bar{I}^{k_{\mathrm{y}}} = 0$).
Also, since the system is now open, $k_{\mathrm{max}}$ is infinite.

Imposing the constraints on the expression of the steady state \eqref{expr:ss}, namely
\begin{equation}
	\left\{ 
	\begin{aligned}
		q &= \bar{Z}^1 \frac{ 1 - k_\mathrm{y} \left( 1 - \bar{r} \right)}{ \left( 1 - \bar{r} \right)^2} \\
		Z^{k_\mathrm{y}} &= \bar{Z}^1 (\bar{r})^{k_\mathrm{y}-1} ,
	\end{aligned}
	\right.
	\label{eq:constr:ss:1}
\end{equation}
we can express the variables $\bar{Z}^1$ and $\bar{r}$ numerically as functions of $q$ and $Z^{k_\mathrm{y}}$, and obtain the equilibrium---exponential---distribution as a function of $q$ and $Z^{k_\mathrm{y}}$.

\subsection{Open system: 2 chemostats}
\label{sec:2chemos}

From two chemostats on, the infinite dimension of the system starts to play a role.
As discussed in the previous section, two chemostats are not enough to drive the network towards a non-equilibrium steady state ($s^{\mathrm{Y}} = 2$, $b = 2$ and $a = 0$):
$I^{k_\mathrm{y1}} = 0$ and $I^{k_\mathrm{y2}} = 0$, where $k_\mathrm{y1}$ and $k_\mathrm{y2}$ represent the masses of the two chemostats ($k_\mathrm{y1} < k_\mathrm{y2}$).
Thus, imposing the known values of the chemostat concentrations on the expression \eqref{expr:ss} leads to
\begin{equation}
	\left\{ 
	\begin{aligned}
		Z^{k_\mathrm{y1}} &= \bar{Z}^1 (\bar{r})^{k_\mathrm{y1}-1} \\
		Z^{k_\mathrm{y2}} &= \bar{Z}^1 (\bar{r})^{k_\mathrm{y2}-1} ,
	\end{aligned}
	\right.
	\label{eq:constr:ss:2}
\end{equation}
which only admits physical solutions if $Z^{k_\mathrm{y1}} > Z^{k_\mathrm{y2}}$.
In this case, from \eqref{eq:constr:ss:2} we obtain the equilibrium distribution
\begin{equation}
	\bar{Z}^k = Z^{k_\mathrm{y1}} \left( \frac{Z^{k_\mathrm{y2}}}{Z^{k_\mathrm{y1}}} \right)^{\frac{k-k_\mathrm{y1}}{k_\mathrm{y2} - k_\mathrm{y1}}} ,
	\label{eq:ss:2}
\end{equation}
which is broader the smaller $Z^{k_\mathrm{y1}} - Z^{k_\mathrm{y2}}$ is or the larger $k_\mathrm{y2} - k_\mathrm{y1}$ is.
When $Z^{k_\mathrm{y1}} \le Z^{k_\mathrm{y2}}$ the equilibrium concentration distribution becomes an increasing exponential which cannot be reached.
As a result the system will enter a regime of continuous growth aimed at reaching that state (which we analyze in sec.~\ref{sec:growth}).

\subsection{Open system: 3 chemostats}
\label{sec:3chemos}

\begin{figure}[t]
	\centering
	\subfloat[][]
	{\includegraphics[width=.45\textwidth]{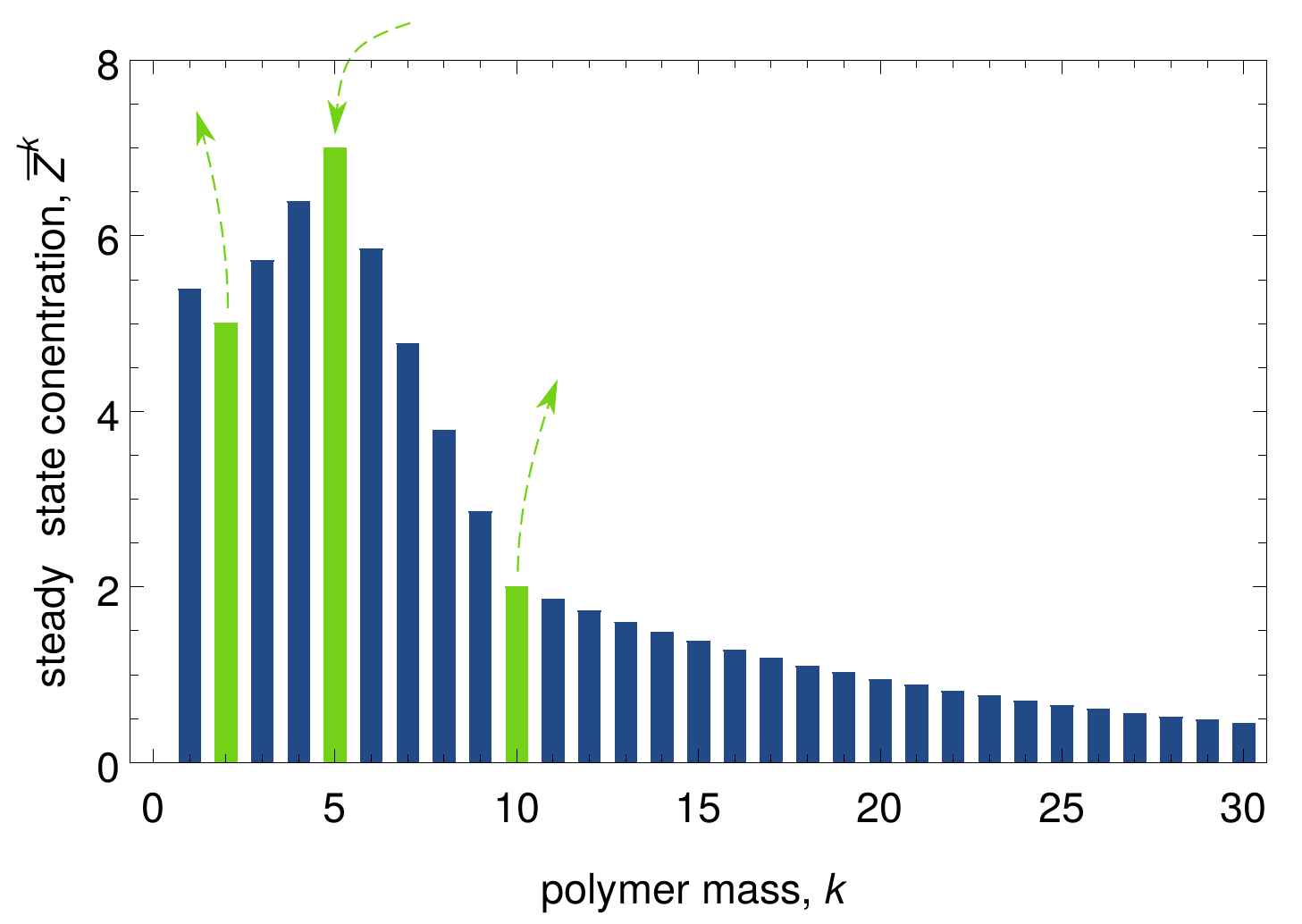} \label{fig:ss3a} } \\
	\subfloat[][]
	{\includegraphics[width=.45\textwidth]{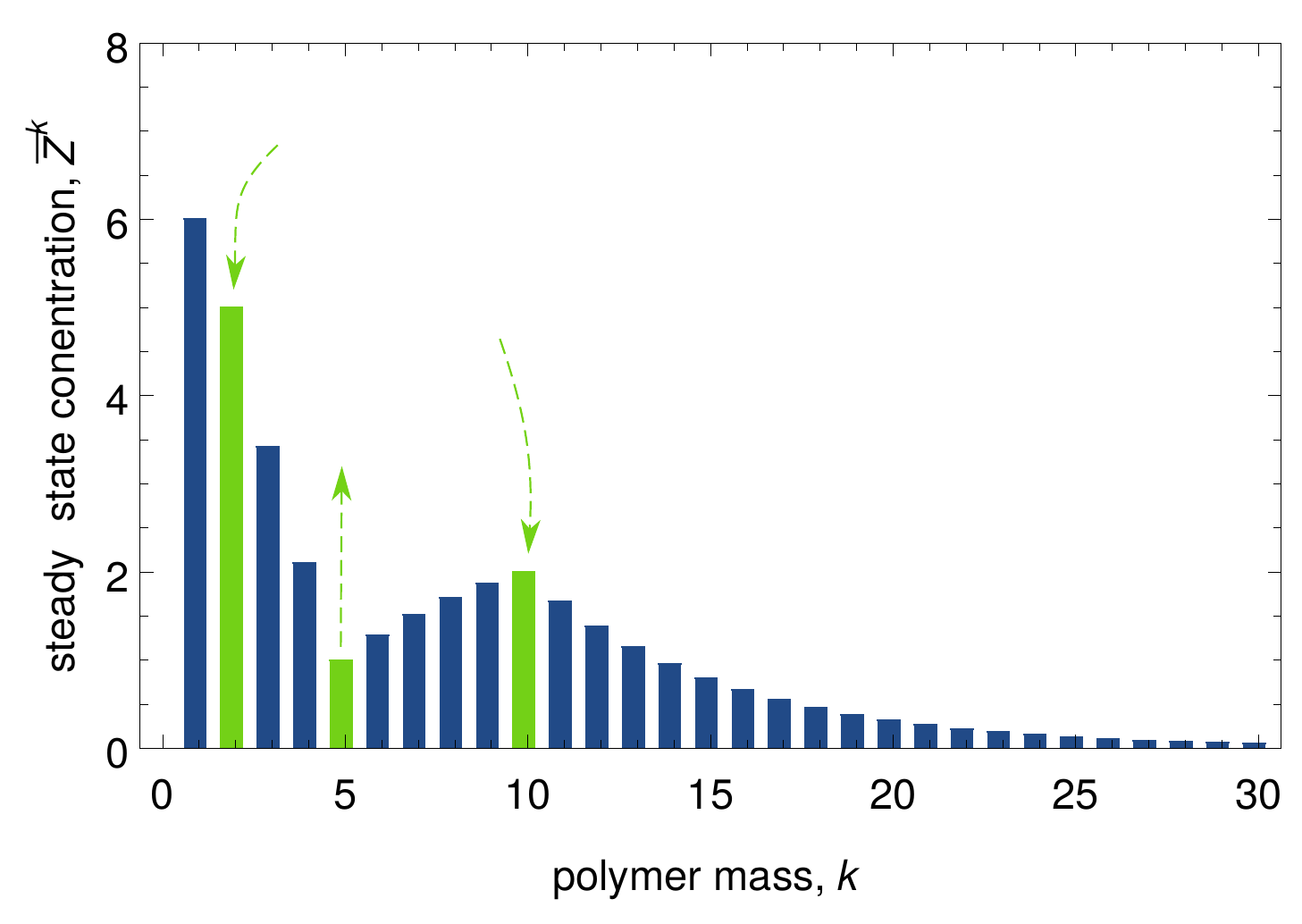} \label{fig:ss3b} }
	\caption{
		Non-equilibrium steady-state distributions for the system of monomer-exchanging polymers with three chemostatted species.
		In both of the plots, the chemostats---highlighted in green and by the arrows---are $k_\mathrm{y1} = 2$, $k_\mathrm{y2} = 5$ and $k_\mathrm{y3} = 10$.
		The orientation of the arrows denotes the sign of the external fluxes of chemostats:
		arrows pointing up means chemostats leaving the system, \emph{i.e.} $I^{k_{\mathrm{y}}} > 0$.
		The chosen chemostat's concentrations are: plot (a) $Z^{k_\mathrm{y1}} = 5$, $Z^{k_\mathrm{y2}} = 7$ and $Z^{k_\mathrm{y3}} = 2$; plot (b) $Z^{k_\mathrm{y1}} = 5$, $Z^{k_\mathrm{y2}} = 1$ and $Z^{k_\mathrm{y3}} = 2$.
	}
	\label{fig:ss3}
\end{figure}

Three is the minimum number of chemostats able to drive the system in a non-equilibrium steady state (sec.~\ref{sec:cycles}).
Indeed, a class of emergent cycles appears ($s^{\mathrm{Y}} = 3$, $b = 2$ and $a = 1$) and the system exhibits a set of non-vanishing external currents.
If we impose the values for the chemostats' concentrations on the general expression for the steady state \eqref{expr:ss}, we obtain:
\begin{equation}
	\left\{
	\begin{aligned}
		\bar{Z}^{k_\mathrm{y1}} &= \bar{Z}^{1} (\bar{r})^{k_\mathrm{y1}-1} \\
		\bar{Z}^{k_\mathrm{y2}} &= \bar{Z}^{1} (\bar{r})^{k_\mathrm{y2}-1} + \frac{\bar{I}^{k_{\mathrm{y1}}}}{\kappa} \, \frac{1 - ( \bar{r} )^{k_\mathrm{y2} - k_{\mathrm{y1}}}}{1 - \bar{r}} \\
		\bar{Z}^{k_\mathrm{y3}} &= \bar{Z}^{1} (\bar{r})^{k_\mathrm{y3}-1} + \frac{\bar{I}^{k_{\mathrm{y1}}}}{\kappa} \, \frac{1 - ( \bar{r} )^{k_\mathrm{y3} - k_{\mathrm{y1}}}}{1 - \bar{r}} + \\ 
		& + \frac{\bar{I}^{k_{\mathrm{y2}}}}{\kappa} \, \frac{1 - ( \bar{r} )^{k_\mathrm{y3} - k_{\mathrm{y2}}}}{1 - \bar{r}} .
	\end{aligned}
	\right.
	\label{eq:constr:ss:3}
\end{equation}
As discussed in sec.~\ref{sec:dissip}, the external currents $\bar{I}^{k_{\mathrm{y}}}$ are subject to the same constraints as the emergent cycles, and can be written as linear combinations of them.
Since we have one class of emergent cycles, characterized by the $\nu^{k_\mathrm{y}}$ values in \eqref{expr:nu3}, we have that
\begin{equation}
	\bar{I}^{k_{\mathrm{y}i}} = \bar{I} \nu^{k_{\mathrm{y}i}}, \quad i = 1,2,3 \;,
	\label{expr:fluxes:3}
\end{equation}
where $\bar{I} \in \R$ determines the exact value of the fluxes.
As for two chemostats, the set of equations in \eqref{eq:constr:ss:3}, in the variables $\bar{Z}^1$, $\bar{r}$ and $\bar{I}$, does not exhibit physical solutions if the concentration of the largest chemostat is higher than the one of the smallest one, i.e. $Z^{k_\mathrm{y1}} \le Z^{k_\mathrm{y3}}$.
On the other hand, whenever the above condition is not fulfilled, the stationary solution is unique and stable (appendix~\ref{app:ss3}).
Solving the system \eqref{eq:constr:ss:3} numerically, we obtain the values of $\bar{Z}^1$, $\bar{r}$ and $\bar{I}$ given $Z^{k_\mathrm{y1}}$, $Z^{k_\mathrm{y2}}$ and $Z^{k_\mathrm{y3}}$.
In Fig.~\ref{fig:ss3} the distribution is shown for different values of these concentrations.

The chemostat concentrations also determine the sign of the related fluxes:
if the concentration of the second chemostat lies above the equilibrium distribution obtained by the first and third one, we have a continuous flow of mass from the intermediate chemostat towards the external ones ($\bar{I} > 0$, Fig.~\ref{fig:ss3a}).
\emph{Vice versa}, if the concentration of the second chemostat lies below the equilibrium distribution obtained by the first and the third one, we have a continuous flow of mass from the smallest and largest chemostats towards the intermediate one ($\bar{I} < 0$, Fig.~\ref{fig:ss3b}).
Importantly, whatever physical value $Z^{k_\mathrm{y1}}$, $Z^{k_\mathrm{y2}}$ and $Z^{k_\mathrm{y3}}$ assume, the system cannot exhibit a condition in which a net flux of matter from the largest species to the smallest one occurs.
This is clear by looking at the $\nu^{k_{\mathrm{y}}}$-values in \eqref{expr:nu3} used to express $\bar{I}^{k_{\mathrm{y}i}}$, eq.~\eqref{expr:fluxes:3}: the sign of $\nu^{k_\mathrm{y1}}$ and $\nu^{k_\mathrm{y3}}$ are always the same, and opposite to the one of $\nu^{k_\mathrm{y2}}$.

\subsection{Open system: more chemostats}
\label{sec:mchemos}

Going on adding chemostats, new independent classes of emergent cycles appear.
The procedure for determining the steady-state distribution is equivalent to that discussed is subsec.~\ref{sec:2chemos} and \ref{sec:3chemos}.
In these two cases we proved that when the largest chemostat has a concentration greater or equal to that of the smallest one, the system does not reach a steady state.
The same exact behavior has been observed numerically for more chemostats, hence we speculate that this property holds for an arbitrary number of chemostats.

\phantom{R.s.}

As a final remark, we point out that the steady-state distributions do not depend on the value of the rate constant $\kappa$.
Indeed, solving the equations \eqref{eq:constr:ss:1}, \eqref{eq:constr:ss:2}, and \eqref{eq:constr:ss:3} for $\bar{Z}^{1}$, $\bar{r}$ and $\bar{I}^{k_{\mathrm{y}}}/\kappa$, we obtain them as functions of the components and the chemostats' concentrations.
Since the latter do not depend on $\kappa$, the same holds for $\bar{Z}^{1}$, $\bar{r}$ and $\bar{I}^{k_{\mathrm{y}}}/\kappa$.
As a corollary $\bar{I}^{k_{\mathrm{y}}}$ is proportional to $\kappa$ and the same holds true for the entropy production \eqref{expr:epr}.

\section{Asymptotic growth regime}
\label{sec:growth}

We mentioned in the previous section that the system does not exhibit a steady state when the concentration of the largest chemostat exceeds that of the smallest one, $Z^{k_{\mathrm{y1}}} \leq Z^{k_{\mathrm{y\,last}}}$---we refer in the text to this configuration of chemostats leading to continuous growth as ``unbalanced''.
The dynamical fixed point moves outside the region of physical solutions---namely to $\bar{r} \geq 1$, see appendix~\ref{app:ss3}---and the system approaches the limit $\bar{r} \rightarrow 1$.
This indicates that the concentration of the single monomer species becomes negligible compared to the rest of the species. 
Hence the system grows towards an unreachable steady state with an exponentially increasing concentration distribution.

\begin{figure}[t]
	\centering
	\subfloat[][]
	{\includegraphics[width=.45\textwidth]{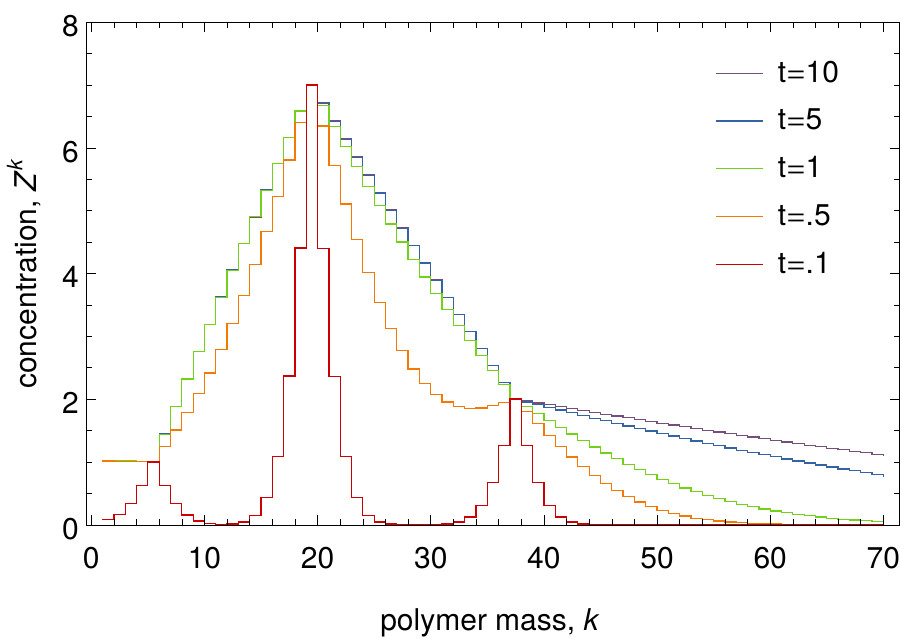} \label{fig:growth3} } \\
	\subfloat[][]
	{\includegraphics[width=.45\textwidth]{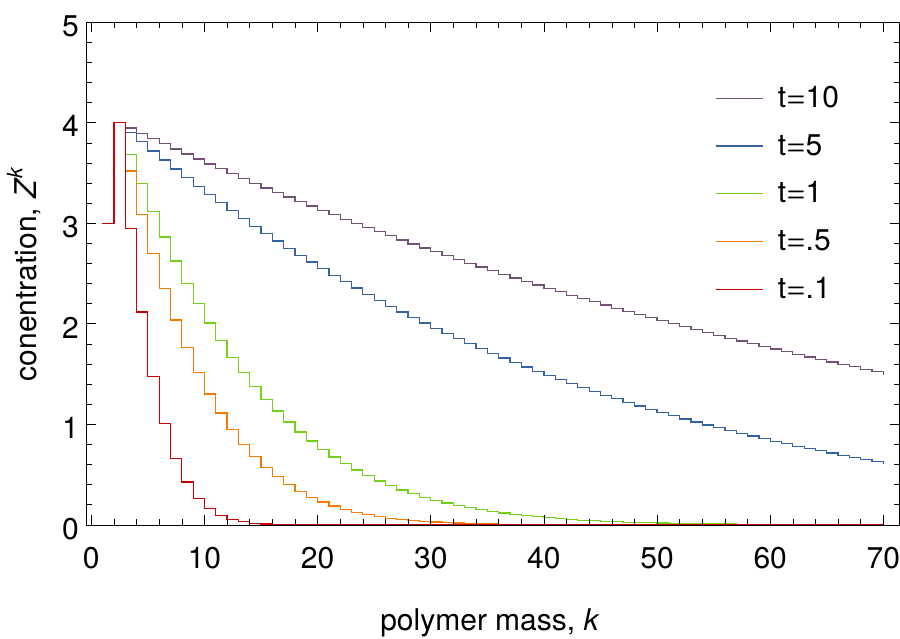} \label{fig:growth2} }
	\caption{
		Concentration distributions at different times are shown for system in unbalanced conditions.
		Different colors from red to violet correspond to exponentially increasing times.
		The set of plots is obtained by numerical solution of the differential equation \eqref{eq:rate:explicit}.
		Absorbing boundary conditions have been chosen, meaning that the concentration at the cut-off---here set to $k_{\mathrm{cut-off}} = 1000$---is zero.
		We point out that this prescription is safe before the cut-off is reached.
		In plot (a) we report a system with three chemostats.
		The chemostat's masses and the related concentrations chosen are: $Z^{5} = 1$, $Z^{19} = 7$ and $Z^{37} = 2$.
		The concentrations of the species between the chemostats basically overlap at times $t \gtrsim 1$ and become steady.
		Beyond this time the growth only involves the species larger than the biggest chemostats.
		In plot (b) we consider a system with monomers and dimers chemostatted: $Z^{1} = 3$ and $Z^{2} = 4$.
	}
	\label{fig:growth}
\end{figure}

Fig.~\ref{fig:growth3} shows the concentration distributions of an unbalanced system at different times before the numerical cut-off (more details are given in the related caption) is reached.
These different distributions show that while the concentrations of the species between two chemostats stabilize to steady values, the concentrations of the species larger than the biggest chemostat do not.
Hence, the system continuously grows trying to populate the infinite size polymer.
This behavior has been observed taking into account different number of chemostats and chemostat' concentrations.

\begin{figure}[t]
\centering
	\subfloat[][balanced chemostats]
		{\includegraphics[width=.33\textwidth]{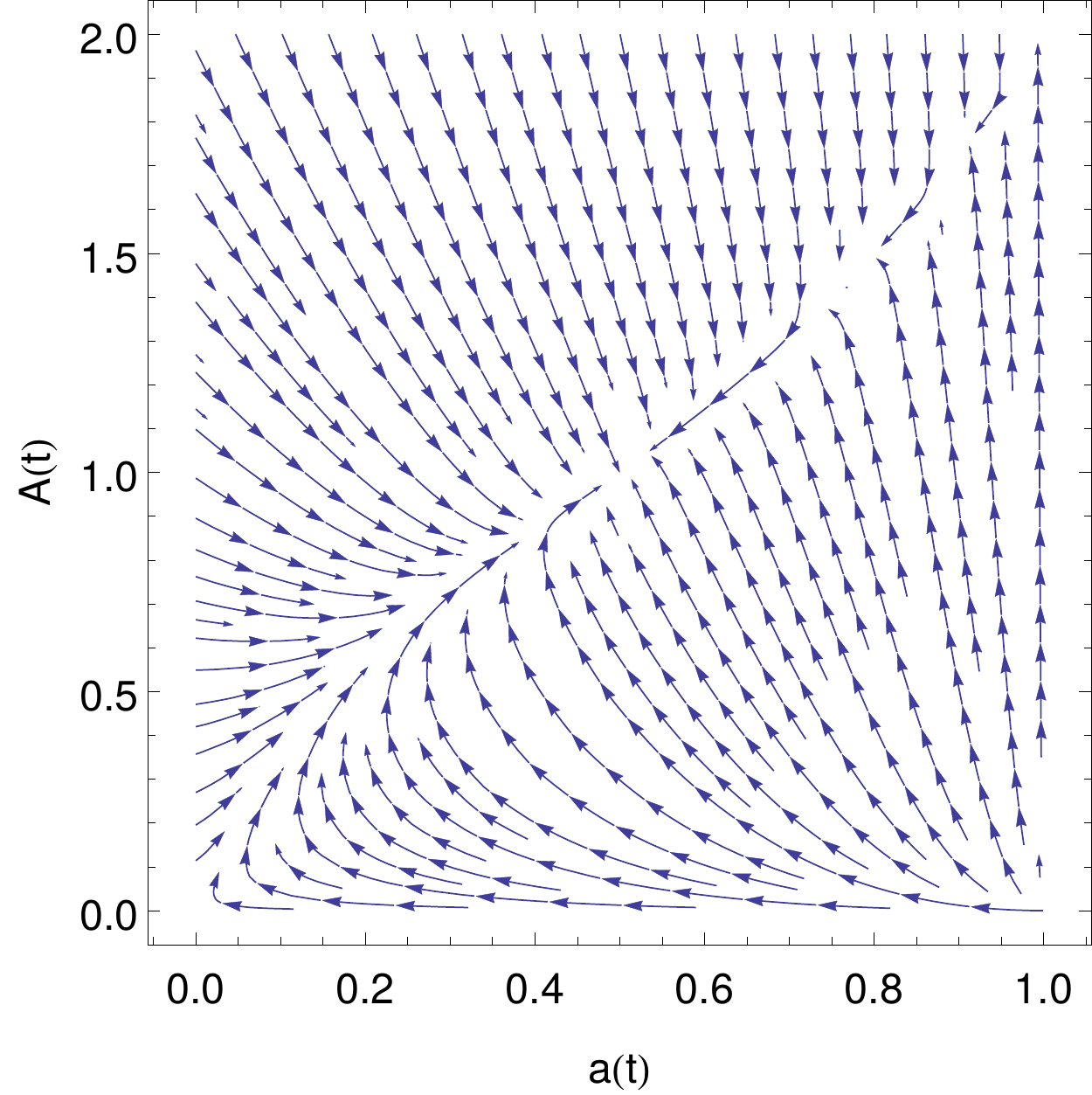} \label{fig:balanced}} \\
	\subfloat[][unbalanced chemostats]
		{\includegraphics[width=.33\textwidth]{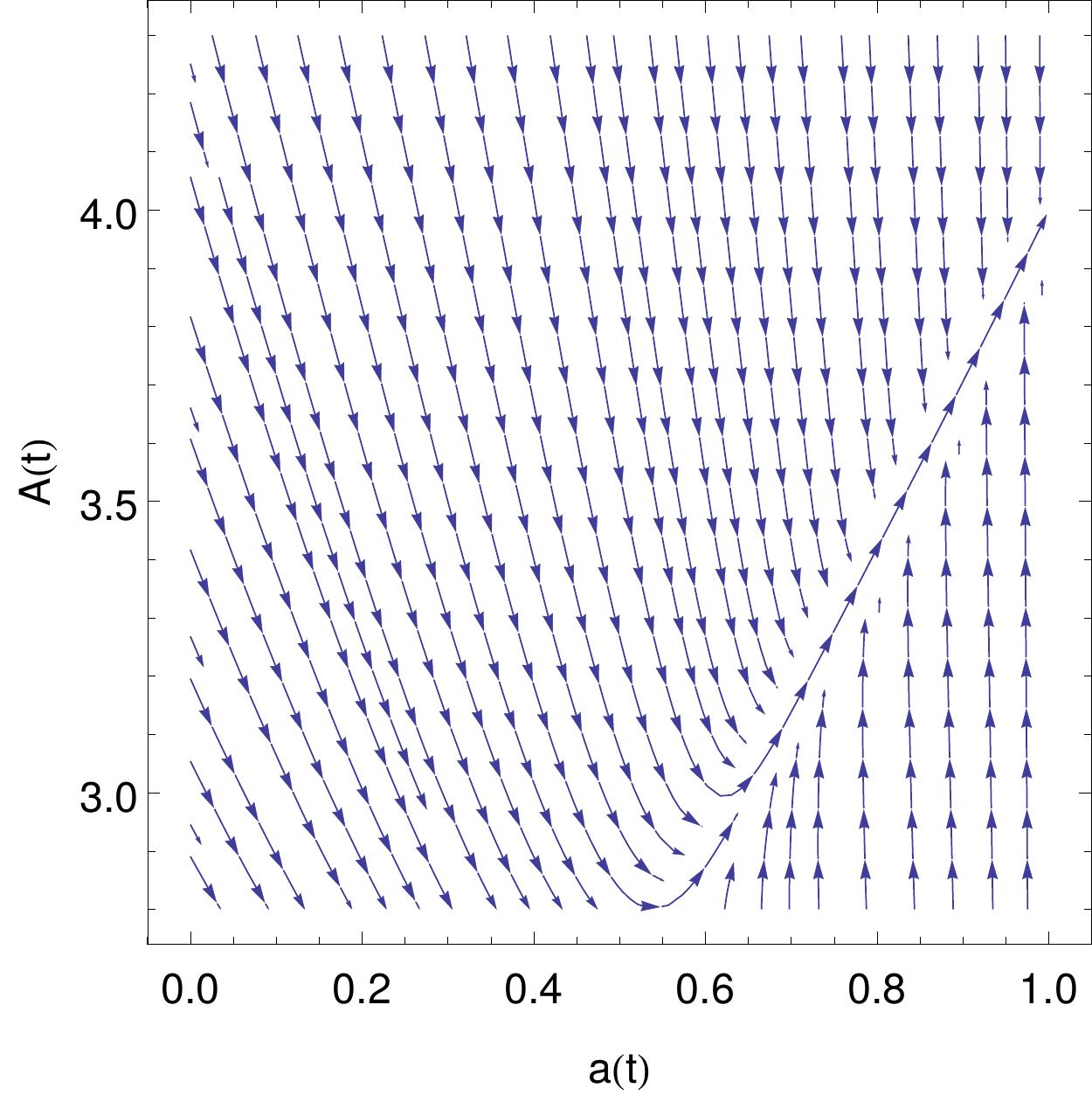} \label{fig:unbalanced}}
	\caption{
		Stream plot of the differential equations \eqref{eq:ansatz} expressed in terms of the ansatz functions $a(t)$ (abscissa) and $A(t)$ (ordinate).
		When balanced chemostat concentrations are used, the fixed point lies for values of $a(t)$ in $]0,1[$: plot (a).
		The chemostats chosen for this plot are $Y^{1} = 4$ and $Y^{2} = 2$.
		\emph{Vice versa}, when the chemostats are unbalanced ($Y^{1} = 2$ and $Y^{2} = 4$) the fixed point moves outside from the physical region ($a(t) > 1$): plot (b).
	}
\label{fig:stream}
\end{figure}

In order to characterize this growth algebraically, we consider a system with monomer and dimer chemostats ($k_\mathrm{y1} = 1$ and $k_\mathrm{y2} = 2$) such that $Z^{k_{\mathrm{y1}}} \leq Z^{k_{\mathrm{y2}}}$.
(The typical growth obtained numerically in this scenario is shown in Fig.~\ref{fig:growth2}).
Since the growth dynamics cannot be solved exactly, we assume that the asymptotic concentration distribution can be parametrized by the (equilibrium) steady state expression \eqref{expr:ss} with time dependent parameters, i.e.
\begin{equation}
	Z^{k}(t) \simeq A(t) \big( a(t) \big)^{k-3}, \quad \text{for } k \geq 3.
	\label{expr:ansatz}
\end{equation}
where $A(t)$ and $a(t)$ are unknown real functions of time.
To simplify the notation, let us denote the concentrations of the chemostats by $Y^1 \equiv Z^{k_\mathrm{y1}}$ and $Y^2 \equiv Z^{k_\mathrm{y2}}$.
The functions $A(t)$ and $a(t)$ can be determined by means of the differential equations for the total concentration $Z$ and the total mass $M$:
\begin{equation}
	\begin{aligned}
		\dot{Z} &= - I^{1} - I^{2} = - \kappa Z ( Z^3 - Y^2 ) - \kappa Y^2 Y^1 \\
		\dot{M} &= - I^{1} - 2 I^{2} = \\
		& - \kappa Z ( 2 Z^3 - 3 Y^2 + Y^1 ) - \kappa 2 Y^2 Y^1 + \kappa Y^1 Y^1 , 
	\end{aligned}
	\label{eq:ansatz}
\end{equation}
where, $Z$, $M$ and the concentration of trimers $Z^3$ assume the following form when the ansatz \eqref{expr:ansatz} is taken into account
\begin{equation}
	\begin{aligned}
		Z(t) &\simeq \frac{A(t)}{1-a(t)}+Y^1+Y^2, \\
		M(t) &\simeq \frac{ 3-2 a(t) }{ \big( 1-a(t) \big)^2} A(t) + Y^1 + 2 Y^2, \\
		Z^3(t) &\simeq A(t) .
	\end{aligned}
	\label{expr:ZMZ3}
\end{equation}
When the equations are expressed in terms of $A(t)$ and $a(t)$, the stream plots for different values of the chemostats' concentrations show that the ansatz captures the non-equilibrium phase transition occurring when the chemostats become unbalanced, Fig.~\ref{fig:stream}.
Indeed, for balanced chemostats, the system evolves towards a fixed point with $a$ lying in $]0,1[$, Fig.~\ref{fig:balanced}.
On the other hand, when the chemostats are unbalanced the fixed point lies beyond $a = 1$ signaling an asymptotic growth regime, see Fig.~\ref{fig:unbalanced}.

\begin{figure}[t]
\centering
	\subfloat[][]
		{\includegraphics[width=.45\textwidth]{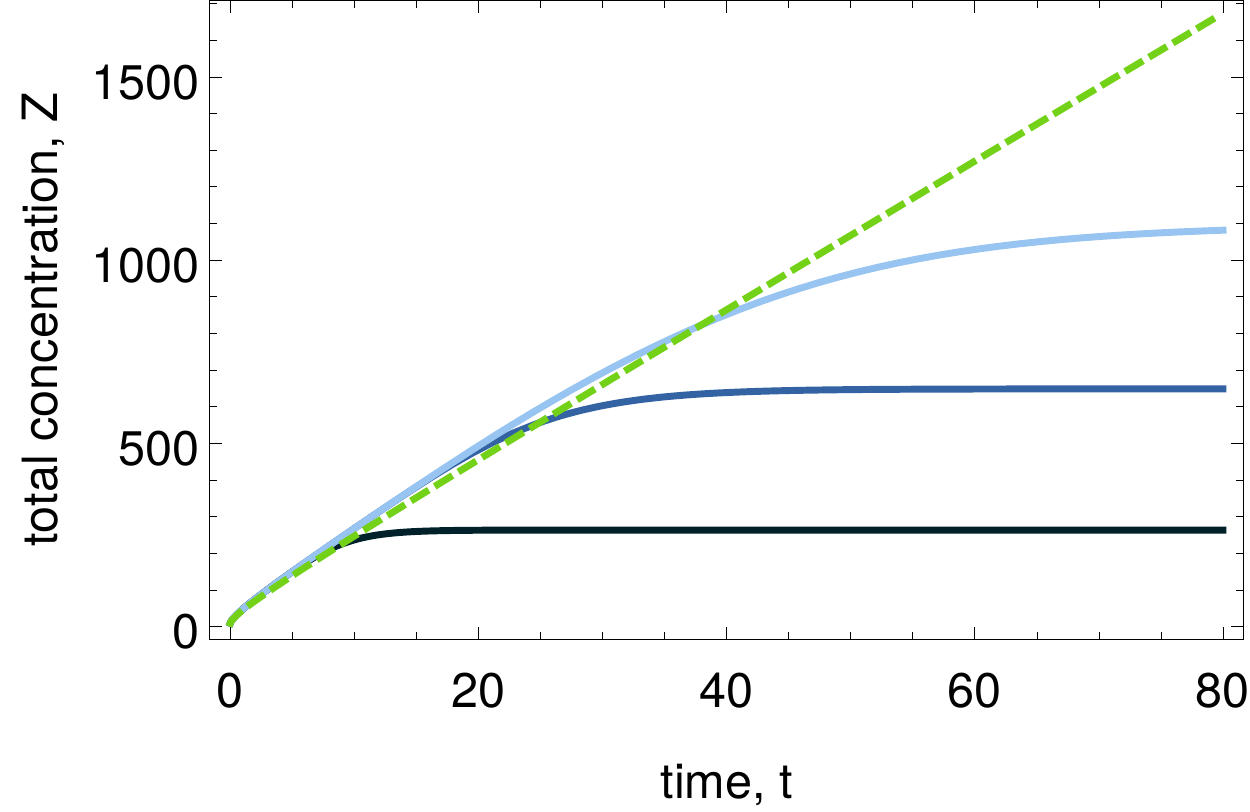}} \\
	\subfloat[][]
		{\includegraphics[width=.45\textwidth]{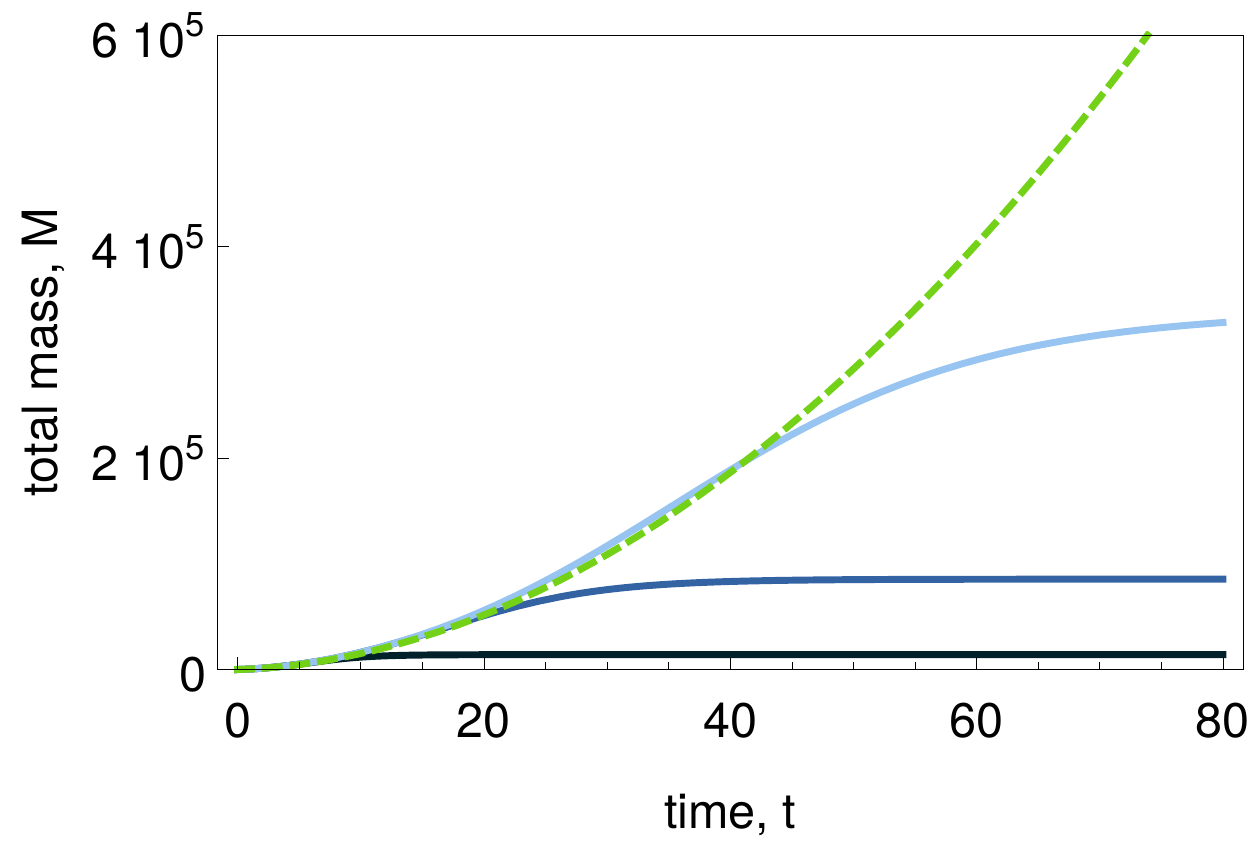}}
	\caption{
		Total concentration (a) and total mass (b) as functions of time in the asymptotic growth regime.
		The numerical solution obtained using the ansatz \eqref{expr:ansatz} is plotted in green (dashed).
		These plots are compared with numerical solutions of the system of differential equations \eqref{eq:rate:explicit} with different cut-offs (blue curves).
		The chosen chemostat concentrations are: $Y^{1} = 3$ and $Y^{2} = 4$ while the initial condition imposed is $Z^{k}(t=0) = \frac{2}{5} (\frac{2}{5})^{k}$.
		Finally, the chosen cut-off concentrations are: $k_{\mathrm{c}} =  200$ (dark blue curve), $k_{\mathrm{c}} =  500$ (blue curve) and $k_{\mathrm{c}} =  1000$ (light blue curve).
	}
\label{fig:ZMgrowth}
\end{figure}

The numerical solution for $A(t)$ and $a(t)$ obtained using \eqref{eq:ansatz} and \eqref{expr:ZMZ3} accurately characterizes the asymptotic growth.
Indeed, as seen in Fig.~\ref{fig:ZMgrowth}, when comparing the evolution of $Z$ and $M$ obtained from $A(t)$ and $a(t)$ with that obtained by solving numerically the rate equations, the former solution overlaps with the latter before the cut-off used in the numerics is reached.
We find that the total concentration grows linearly with time whereas the mass quadratically.

\begin{figure}[t]
\centering
	\includegraphics[width=.45\textwidth]{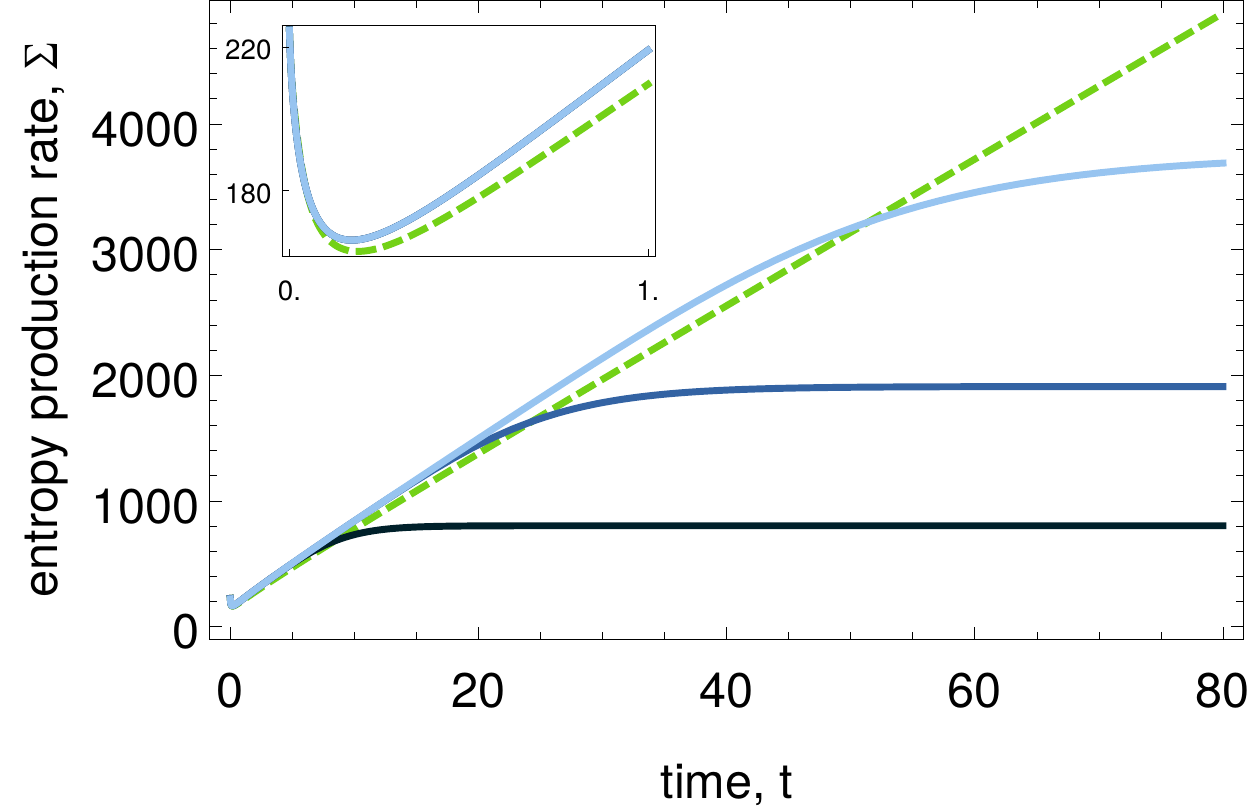}
	\caption{
		Entropy production rate as a function of time in the asymptotic growth regime.
		The numerical solution obtained using the ansatz \eqref{expr:ansatz} is plotted in green (dashed).
		This plot is compared with numerical solutions of the system of differential equations \eqref{eq:rate:explicit} with different cut-offs (blue curves).
		In all the plot, the entropy production rate is given in units of $R$.
		The chosen chemostat concentrations are: $Y^{1} = 3$ and $Y^{2} = 4$ while the initial condition imposed is $Z^{k}(t=0) = \frac{2}{5} (\frac{2}{5})^{k}$.
		The chosen cut-offs $k_{\mathrm{c}}$ are: 200 (dark blue curve), 500 (blue curve) and 1000 (light blue curve).
		Also, the inset shows in greater details the initial transient relaxation stage.
	}
\label{fig:epr}
\end{figure}

Taking into account the ansatz \eqref{expr:ansatz}, the entropy production rate \eqref{expr:epr} becomes
\begin{equation}
	\Sigma \simeq R I^{1} \ln \frac{A(t)}{Y^{1} (a(t))^{2}} + R I^{2} \ln \frac{A(t)}{Y^{2}a(t)} \, ,
\end{equation}
where, $I^{1}$ and $I^{2}$ can be written in terms of $Y^{1}$, $Y^{2}$, $A(t)$ and $a(t)$ using eq.~\eqref{eq:ansatz}.
The latter is plotted in Fig.~\ref{fig:epr}, where it is compared with the numerical solutions for different cut-offs.
The agreement with the numerical solution is not perfect but captures the linear asymptotic growth of the entropy production rate reasonably well.
Also, we point out that the unbalanced dynamics shown in Fig.~\ref{fig:epr} exhibits an initial transient relaxation stage shown in inset.

We conclude mentioning that the same ansatz could be used for systems characterized by more chemostats with unbalanced concentrations.
Indeed, the growth always involves the species larger than the biggest chemostat, whereas the species between chemostats converge faster to proper steady values.
Hence, fixing the concentration of these latter species, we could assume a growth like \eqref{expr:ansatz} for the species larger then the biggest chemostat and perform the same analysis.

\section{Conclusions}
\label{sec:conc}

This paper provides a kinetic description of systems made of glucans and processed by the class of enzymes known as D-enzymes.
The action of the enzyme induces a monomer-exchange process \cite{krapivsky10} between pairs of glucans which are distinguished by their mass or degree of polymerization. 
Free monomers are not allowed to attach to other polymers \cite{jones69} implying that the total concentration and the total mass are conserved when the system is closed.
The system's dynamics is ruled by rate equations for the polymer concentrations endowed with mass action kinetics.
We mimic physiological conditions by introducing \emph{chemostats} which effectively describe the action of the environment by fixing the concentrations of certain glucans.
	In this scenario, chemostats represent species processed by the environment. For example, they may represent species which need to be processed and injected by the environment in the system, analogously, they may represent final products of the metabolic processes which are taken out of the system.
Importantly, chemostatting the system amounts to open it and introduce driving forces on the non-chemostatted species.  

Our main results are summarized in Table~\ref{tab:summary}.
We identified three types of different long-time behaviors depending on the chemostatting conditions: equilibrium state, non-equilibrium steady state, and continuous growth of the system.
The closed system as well as the open system with a single chemostat always relax to an equilibrium state.
In presence of two chemostats the system will either relax to equilibrium or turn into a state of continuous growth depending on whether or not the concentration of the largest chemostat is lower than the concentration of the smallest one.
We proved that this latter condition for growth holds true for up to three chemostats and conjectured that it is generally true based on numerical evidence.
For more than two chemostats, if the concentration of the largest chemostat is lower than that of the smallest one, the system will reach a nonequilibrium steady state where the chemostats continuously exchanges matter across the system.
	Our results confirm that, even in the infinite-dimensional chemical network considered here, the number of chemostats equals to the number of broken conservation laws plus the number of emergent cycles (see Tab.~\ref{tab:summary}).
	A proof of this equality for finite dimensional chemical networks is provided in Ref.~\cite{polettini14}.
	We also emphasized the role of the emergent cycles in driving the chemostatted chemical networks towards nonequilibrium steady states rather than equilibrium states \cite{polettini14}.

The metabolism of polysaccharides is a complex process involving many steps and several enzymes \cite{ball03} and its complete dynamical characterization is beyond the scope of the present paper.
We focused on the dynamical characterization of the disproportionating action of D-enzymes in the breakdown and synthesis processes of glucans \cite{takaha99}.
Under physiological conditions, it has been pointed out that one of the possible role of D-enzymes in these processes is to produce glucans of large sizes (which are then processed by other enzymes) starting from medium sized ones \cite{takaha99}.
	Importantly, a production of glucose (monomers in our descriptions) is expected, too \cite{takaha99}.
	This disproportionating behavior can be reproduced in a (nonequilibrium) steady state by the three chemostats system depicted in Figure \ref{fig:ss3a}.
	The intermediate high concentration chemostatted glucans represent the species to be processed, while the low concentration chemostatted glucans represent the species to be produced---in this case the small and large glucans.
	In this scenario, a continuous flow of intermediate glucans enters the system and consequently both the smaller and the larger glucans are steadily produced and expelled from the system (\S~\ref{sec:3chemos}).
	We stress that the production of the small glucans follows from the total concentration conservation law  (\S~\ref{sec:cl}), \emph{i.e.} the fact that free monomers cannot attach to other glucans.
	As seen in sec.~\ref{sec:2chemos}, two chemostats are not sufficient to reproduce a nonequilibrium steady state.

Also, under closed in vitro conditions, the equilibrium distribution (which has also been analyzed in Ref. \cite{lahiri15} and can be equivalently obtained by means of Maximum Entropy methods \cite{kartal11}) agrees with experiments \cite{kartal11}.
This means that if chemostatting conditions could be implemented \emph{in vitro}, our predictions could be verified experimentally.
Such a procedure would also enable to engineer different polymer concentration distributions.

The approach we developed could be easily extended to describe the behavior of more sophisticated forms of D-enzymes \cite{kartal11} embedding further conservation laws.
It is also relevant to study any type of exchange process or aggregation--fragmentation dynamics \cite{krapivsky10} in an open system framework \cite{budrikis14,knowles07,knowles09,cohen13}, emphasizing the importance of conservation laws and providing more insights into the mechanisms driving these processes out of equilibrium.

\begin{acknowledgments}

	R.R. is grateful to A. Wachtel for valuable discussions and suggestions.
	The present project was supported by the National Research Found, Luxembourg, in the frame of project FNR/A11/02 and of the AFR PhD Grant 2014-2, No.~9114110.

\end{acknowledgments}

\appendix

\section{Steady-state distributions}
\label{app:ss}

The generic expression for the steady-state distribution \eqref{expr:ss} can be obtained as follows.
The steady-state equations can be expressed as
\begin{equation}
	\begin{gathered}
		\bar{Z} \left\{ \bar{Z}^2 - \bar{Z}^1 \right\} + \bar{Z}^1 \bar{Z}^1 = 0, \\
		\begin{gathered}
			\bar{Z} \left\{ \bar{Z}^{k+1} - 2 \bar{Z}^k + \bar{Z}^{k-1} \right\} + \bar{Z}^1 \left\{ \bar{Z}^{k} - \bar{Z}^{k-1} \right\} \\
			= \frac{\bar{I}^k}{\kappa} \delta_{k \, k_{\mathrm{y}}\in\Omega_{\mathrm{Y}}} , \quad \text{for } k \geq 2 .
		\end{gathered}
	\end{gathered}
\end{equation}
Defining the variable $\Delta \bar{Z}^{k} \equiv \bar{Z}^{k} - \bar{Z}^{k-1}$ they become
\begin{equation}
	\begin{gathered}
		\bar{Z} \Delta \bar{Z}^2 + \bar{Z}^1 \bar{Z}^1 = 0, \\
		\begin{gathered}
				\bar{Z} \left\{ \Delta \bar{Z}^{k+1} - \Delta \bar{Z}^k \right\} + \bar{Z}^1 \Delta \bar{Z}^{k} \\
					= \frac{\bar{I}^k}{\kappa} \delta_{k \, k_{\mathrm{y}}\in\Omega_{\mathrm{Y}}} , \quad \text{for } k \geq 2 .
		\end{gathered}
	\end{gathered}
\end{equation}
Hence, by hierarchically substituting these expression one into the other, and using the variable $\bar{r} \equiv 1 - {\bar{Z^{1}}}/{\bar{Z}}$, we obtain
\begin{multline}
	\Delta \bar{Z}^{k} = - \left( 1 -  \bar{r} \right) \bar{Z}^1  \bar{r}^{k-2} + \\ + \sum_{k_{\mathrm{y}} \in \Omega_{\mathrm{Y}}} \frac{\bar{I}^{k_{\mathrm{y}}}}{\kappa} \, \bar{r}^{k - k_{\mathrm{y}} - 1} \, \Theta \left( k - k_{\mathrm{y}} - 1 \right) \, ,
\end{multline}
where $\Theta(\cdot)$ represents the discrete step function:
\begin{equation}
	\Theta(k) = 
		\begin{cases}
			0 & \text{if } k < 0 \, , \\
			1 & \text{if } k \ge 0 \, .
		\end{cases}
	\label{ex:theta}
\end{equation}
Finally,
\begin{multline}
	\bar{Z}^{k} = \sum_{i = 1}^{k} \Delta \bar{Z}^{i} = \bar{Z}^{1} (\bar{r})^{k-1} + \\ + \sum_{k_{\mathrm{y}} \in \Omega_{\mathrm{Y}}} \frac{\bar{I}^{k_{\mathrm{y}}}}{\kappa} \, \frac{1 - ( \bar{r} )^{k - k_{\mathrm{y}}}}{1 - \bar{r}} \, \Theta \left( k - k_{\mathrm{y}} - 1 \right) ,
\end{multline}
which corresponds to the equation \eqref{expr:ss} in the main text.

\section{Three chemostats steady state}
\label{app:ss3}

We discuss the uniqueness and stability conditions for the steady state when three chemostats are present.

From the constraints on the steady state \eqref{eq:constr:ss:3} and from the condition for the external currents \eqref{expr:fluxes:3}, we can write a single steady state condition involving just $\bar{r}$ as variable:
\begin{multline}
	\begin{aligned}
		& \left( \nu^{k_{\mathrm{y3}}} \bar{Z}^{k_{\mathrm{y1}}} + \nu^{k_{\mathrm{y1}}} \bar{Z}^{k_{\mathrm{y2}}} \right) (\bar{r})^{ \nu^{k_{\mathrm{y1}}} + \nu^{k_{\mathrm{y3}}}} - \\
		& \left( \nu^{k_{\mathrm{y1}}} + \nu^{k_{\mathrm{y3}}} \right) \bar{Z}^{k_{\mathrm{y2}}} (\bar{r})^{\nu^{k_{\mathrm{y1}}}} - \\
		& \left( \nu^{k_{\mathrm{y3}}} \bar{Z}^{k_{\mathrm{y1}}} + \nu^{k_{\mathrm{y1}}} \bar{Z}^{k_{\mathrm{y3}}} \right)  (\bar{r})^{ \nu^{k_{\mathrm{y3}}} } + \\
		& \left( \nu^{k_{\mathrm{y3}}} \bar{Z}^{k_{\mathrm{y2}}} + \nu^{k_{\mathrm{y1}}} \bar{Z}^{k_{\mathrm{y3}}} \right) = 0 .
	\end{aligned}
\end{multline}
Let us define the variables $x \equiv (\bar{r})^{\nu^{k_{\mathrm{y3}}}}$ and $y \equiv (\bar{r})^{\nu^{k_{\mathrm{y1}}}}$, so that the above-expressed steady-state condition can be written as the intersection of two curves: a rectangular hyperbola and a power law function
\begin{equation}
	\left\{
	\begin{aligned}
		y_{\mathrm{h}} &= y_0 - \frac{z_0}{x-x_0} \\
		y_{\mathrm{p}} &= (x)^{\nu^{k_{\mathrm{y1}}} / \nu^{k_{\mathrm{y3}}}} ,
	\end{aligned}
	\right.
	\label{eq:hp}
\end{equation}
where the coefficients are given by
\begin{equation}
	\begin{aligned}
		x_0 &= \frac{\left( \nu^{k_{\mathrm{y1}}} + \nu^{k_{\mathrm{y3}}} \right) \bar{Z}^{k_{\mathrm{y2}}}}{\nu^{k_{\mathrm{y3}}} \bar{Z}^{k_{\mathrm{y1}}} + \nu^{k_{\mathrm{y1}}} \bar{Z}^{k_{\mathrm{y2}}}} , \\
		y_0 &= \frac{\nu^{k_{\mathrm{y3}}} \bar{Z}^{k_{\mathrm{y1}}} + \nu^{k_{\mathrm{y1}}} \bar{Z}^{k_{\mathrm{y3}}}}{\nu^{k_{\mathrm{y3}}} \bar{Z}^{k_{\mathrm{y1}}} + \nu^{k_{\mathrm{y1}}} \bar{Z}^{k_{\mathrm{y2}}}} , \\
		z_0 &= \frac{ \nu^{k_{\mathrm{y1}}} \nu^{k_{\mathrm{y3}}} \left( \bar{Z}^{k_{\mathrm{y2}}} - \bar{Z}^{k_{\mathrm{y1}}} \right) \left( \bar{Z}^{k_{\mathrm{y2}}} - \bar{Z}^{k_{\mathrm{y3}}} \right)}{ \left( \nu^{k_{\mathrm{y3}}} \bar{Z}^{k_{\mathrm{y1}}} + \nu^{k_{\mathrm{y1}}} \bar{Z}^{k_{\mathrm{y2}}} \right)^2} .
	\end{aligned}
	\label{eq:system}
\end{equation}
[The subscripts $\mathrm{h}$ and $\mathrm{p}$ simply help us to distinguish the two functions.]
From a geometrical point of view, physical solutions are represented by those intersection points lying in $(x,y) \in (0,1) \times (0,1)$.
In order to prove that this happens whenever $\bar{Z}^{k_{\mathrm{y1}}} > \bar{Z}^{k_{\mathrm{y3}}}$ we observe that all of the possible configurations of chemostat concentrations are described by the following four cases for the parameters $x_0$ and $y_0$.

\begin{figure}[t]
\centering
	\subfloat[][$x_0 < 1$ and $y_0 < 1$]
		{\includegraphics[width=.22\textwidth]{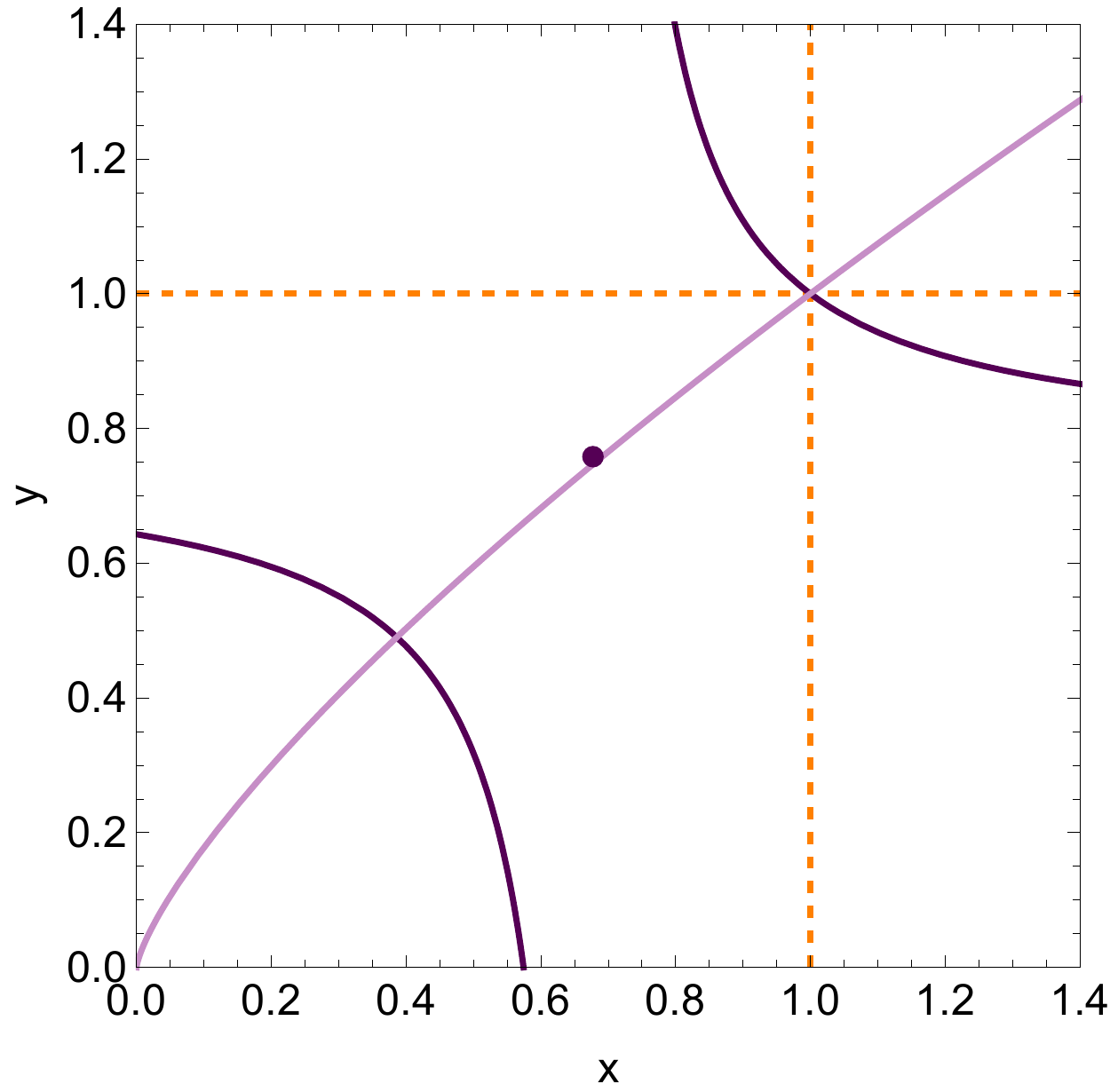} \label{fig:hp1}} \quad
	\subfloat[][$x_0 < 1$ and $y_0 > 1$]
		{\includegraphics[width=.22\textwidth]{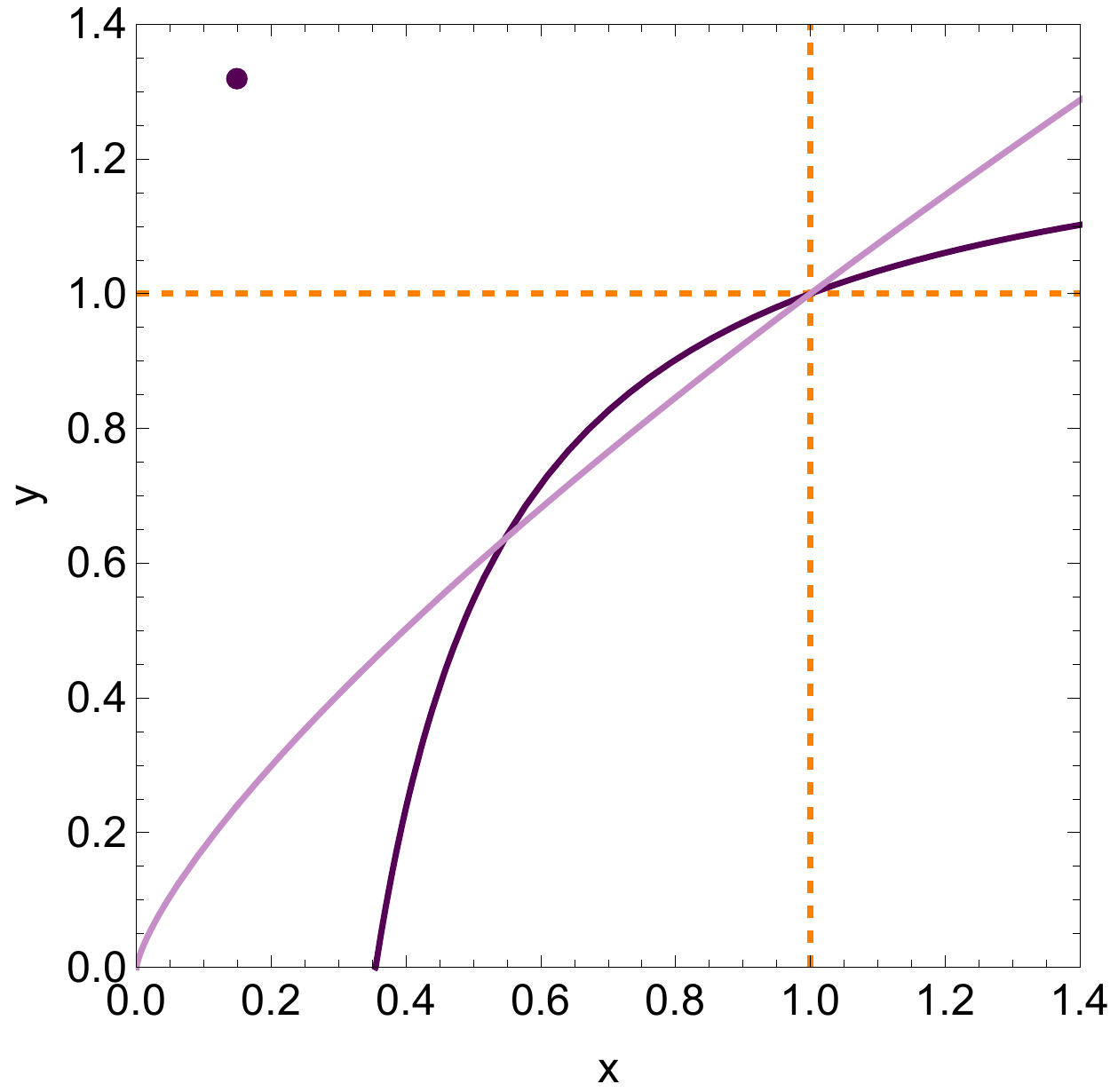} \label{fig:hp2}} \\
	\subfloat[][$x_0 > 1$ and $y_0 < 1$]
		{\includegraphics[width=.22\textwidth]{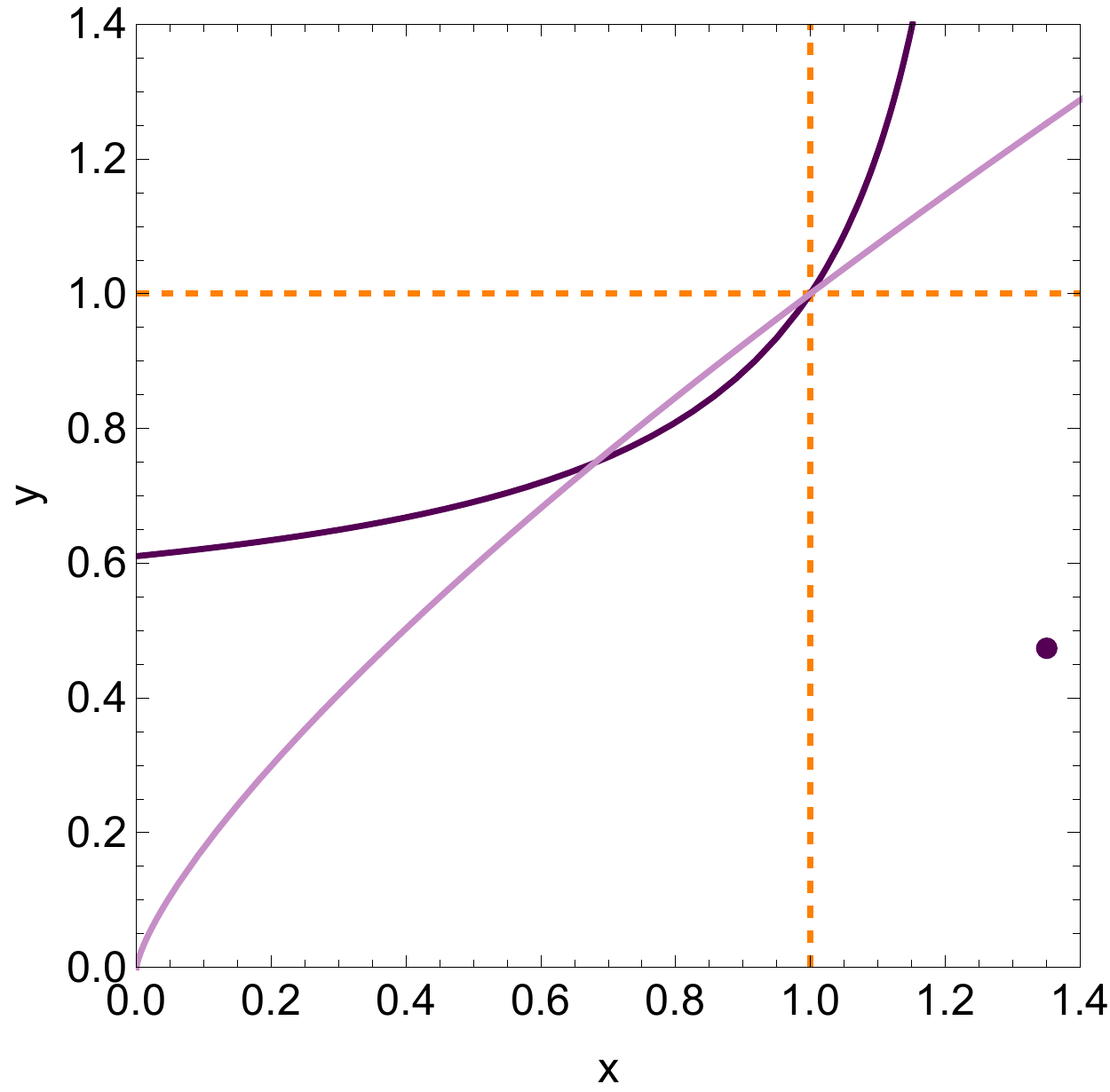} \label{fig:hp3}} \quad
	\subfloat[][$x_0 > 1$ and $y_0 > 1$]
		{\includegraphics[width=.22\textwidth]{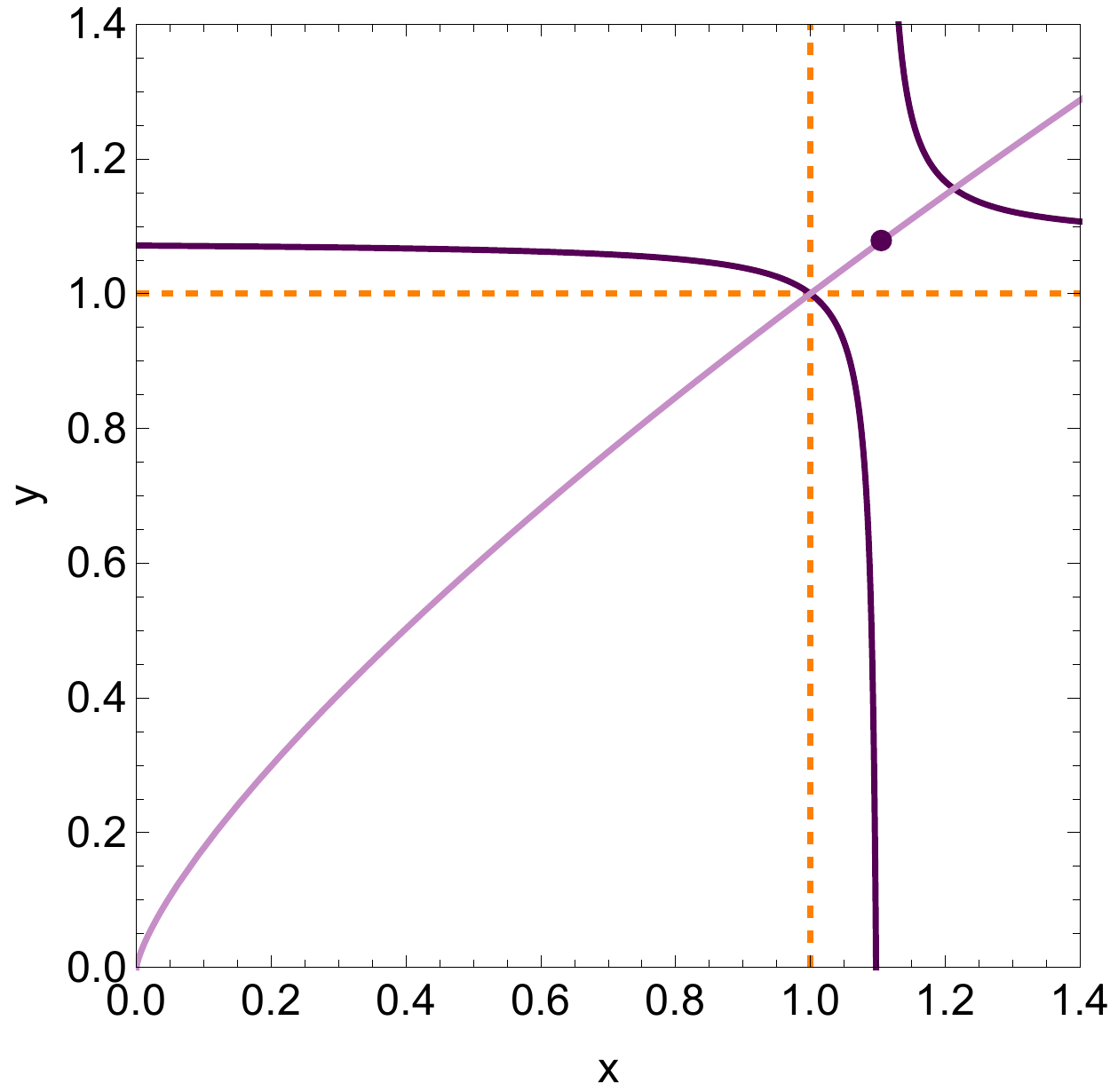} \label{fig:hp4}}
	\caption{ 
		Plots of the hyperbola (dark purple curve) and power law (light purple curve) in \eqref{eq:hp} for different configurations of parameters.
		The center of the hyperbola is highlighted by a dark purple dot, while the physical region by the dashed orange lines.
	}
\label{fig:hp}
\end{figure}

\begin{itemize}
	\item $x_0 < 1$ and $y_0 < 1$ ($z_0 < 0$). 
		This condition implies the following configuration for the chemostats: $\bar{Z}^{k_{\mathrm{y1}}} > \bar{Z}^{k_{\mathrm{y2}}} > \bar{Z}^{k_{\mathrm{y3}}}$.

		In this case we have always one and only one solution.
		Indeed, the center of the hyperbola ($x_0, y_0$) takes $(0,1) \times (0,1)$, and the upper right branch of the hyperbola always intersects the power law in $x = 1$ (which is non-physical).
		The left lower one, instead, always intersects the power law for values in $(0,1) \times (0,1)$ since $y_{\mathrm{h}}(x = 0) > 0$ (fig.~\ref{fig:hp1}).
	\item $x_0 < 1$ and $y_0 > 1$ ($z_0 > 0$). 
		This condition corresponds to $\bar{Z}^{k_{\mathrm{y1}}} > \bar{Z}^{k_{\mathrm{y2}}}$ and $\bar{Z}^{k_{\mathrm{y3}}} > \bar{Z}^{k_{\mathrm{y2}}}$.

		In this case we have one solution if and only if $\bar{Z}^{k_{\mathrm{y1}}} > \bar{Z}^{k_{\mathrm{y3}}}$.
		The center of the hyperbola lies in $(0,1) \times (1,\infty)$ and the upper left branch of the hyperbola never intersects the power law.
		The right lower one, instead, always intersects the power law in $x = 1, \, y = 1$ (fig.~\ref{fig:hp2}).
		We have a further intersection in the physical region if and only if $\at{\der{y_{\mathrm{p}}}{x}}{x = 1} > \at{\der{y_{\mathrm{p}}}{x}}{x = 1}$, which holds iff $\bar{Z}^{k_{\mathrm{y1}}} > \bar{Z}^{k_{\mathrm{y3}}}$---indeed, $x^{\ast}:\,y_{\mathrm{h}}(x^{\ast}) = 0$ is such that $x^{\ast} > 0$, for any choice of the chemostats. 
	\item 
$x_0 > 1$ and $y_0 < 1$ ($z_0 > 0$).
This condition corresponds to: $\bar{Z}^{k_{\mathrm{y1}}} < \bar{Z}^{k_{\mathrm{y2}}}$ and $\bar{Z}^{k_{\mathrm{y3}}} < \bar{Z}^{k_{\mathrm{y2}}}$.

Once again, we have one solution if and only if $\bar{Z}^{k_{\mathrm{y1}}} > \bar{Z}^{k_{\mathrm{y3}}}$.
The center of the hyperbola lies in $(1,\infty) \times (0,1)$ and the right lower branch of the hyperbola never intersects the power law.
The upper left one, instead, always intersects the power law in $x = 1, \, y = 1$ (fig.~\ref{fig:hp3}).
We have a further intersection in the physical region if and only if $\at{\der{y_{\mathrm{p}}}{x}}{x = 1} > \at{\der{y_{\mathrm{p}}}{x}}{x = 1}$, which holds iff $\bar{Z}^{k_{\mathrm{y1}}} > \bar{Z}^{k_{\mathrm{y3}}}$---indeed, $y_{\mathrm{h}}(0) > 0$ for any choice of the chemostats.
	\item 
$x_0 > 1$ and $y_0 > 1$ ($z_0 < 0$).
This condition implies the following configuration for the chemostats: $\bar{Z}^{k_{\mathrm{y1}}} < \bar{Z}^{k_{\mathrm{y2}}} < \bar{Z}^{k_{\mathrm{y3}}}$.

In this case we have no solutions.
Indeed, the center of the hyperbola lies in $(x,y) \in (1,\infty) \times (1,\infty)$ and neither the upper right nor the lower left branch of the hyperbola intersects the power law in the physical region.
The left lower one, indeed, always intersects the power law in $(1,1)$ which is non-physical (fig.~\ref{fig:hp3}).
\end{itemize}

Summarizing, we have a unique steady state whenever the concentration of the largest chemostat is higher than the concentration of the smallest one: $\bar{Z}^{k_{\mathrm{y1}}} > \bar{Z}^{k_{\mathrm{y3}}}$.

\emph{Stability.} In order to prove the stability of the fixed point we resort to the following Lyapunov function:
\begin{equation}
	L = \sum_{k} Z^{k} \ln \frac{Z^{k}}{Z^{k}_{\mathrm{s}}} - \left( Z - Z_{\mathrm{s}} \right) .
	\label{expr:lyapunov}
\end{equation}
It is easy to prove that this function is always positive and vanishes only for $Z^{k} = Z^{k}_{\mathrm{s}}$, where $Z^{k}_{\mathrm{s}}$ represents the steady-state solution.
If the steady-state solution exists, namely if exists $Z^{k}_{\mathrm{s}} : \dot{Z^{k}_{\mathrm{s}}} = 0$, the time derivative of the Lyapunov function \eqref{expr:lyapunov} can be written as
\begin{equation}
	\der{L}{t} = \sum_{k_{\mathrm{x}}} \dot{Z}^{k_{\mathrm{x}}} \ln \frac{Z^{k_{\mathrm{x}}}}{Z^{k_{\mathrm{x}}}_{\mathrm{s}}} .
	\label{expr:lyder}
\end{equation}
Close to the steady state the above derivative is negative.
Spanning the phase space with small perturbations on every concentration, we always obtain $\der{L}{t} \le 0$, where the equal sign is reached only at the steady state.
Disregarding the infinite dimension of the phase space, we consider the independent set of perturbations labeled with the index $k'_{\mathrm{x}}$ and quantified by the small real value $\epsilon$
\begin{equation}
	Z^{k_{\mathrm{x}}} = Z^{k_{\mathrm{x}}}_{\mathrm{s}} + \epsilon \delta^{k'_{\mathrm{x}} k_{\mathrm{x}}} , \quad | \epsilon | \ll \min_{k_{\mathrm{x}}} Z^{k_\mathrm{x}}_{\mathrm{s}} .
	\label{expr:close}
\end{equation}
Embedding these perturbation in \eqref{expr:lyder} and using the rate equations \eqref{eq:rate:explicit} we obtain
\begin{equation}
	\begin{aligned}
		\der{L}{t} & \simeq - \frac{\kappa}{Z^1_\mathrm{s}} \left( Z_\mathrm{s} - Z^1_\mathrm{s} - Z^2_\mathrm{s} \right) \epsilon^2 , \quad \text{for } k'_\mathrm{x} = 1, \\
		\der{L}{t} & \simeq - \frac{\kappa}{Z^{k'_\mathrm{x}}_\mathrm{s}} \left( 2 Z_\mathrm{s} + 2 Z^{k'_\mathrm{x}}_{\mathrm{s}} + \right. \\ 
		& \left. - Z^{k'_\mathrm{x}+1}_\mathrm{s} - Z^{k'_\mathrm{x}-1}_\mathrm{s} - Z^{1}_{\mathrm{s}} \right) \epsilon^2 , \quad \text{for } k'_\mathrm{x} \neq 1 ,
	\end{aligned}
	\label{expr:derapprox}
\end{equation}
which are always negative, no matter the sign of the perturbation.

\bibliography{denzymesBib,BibFile}

\end{document}